\documentclass[
 reprint,
 superscriptaddress,shortbibliography,
 amsmath,amssymb,
 aps,
 prl,
 twocolumn,
nofootinbib
floatfix,
]{revtex4-2}
\usepackage{color,soul}
\usepackage{physics}
\usepackage{graphicx}
\usepackage[utf8]{inputenc}
\usepackage{graphicx}
\usepackage{dcolumn}
\usepackage{graphicx}
\usepackage{amsmath}
\usepackage{caption}
\usepackage{subcaption}
\usepackage{array}
\usepackage{wrapfig}
\usepackage{booktabs}
\usepackage{esint}
\usepackage{hyperref}
\usepackage{adjustbox}
\usepackage[most]{tcolorbox}
\usepackage{caption}
\usepackage{subcaption}
\usepackage{tikz}
\usepackage{pgfplots}
\usepackage{ragged2e}
\usepgfplotslibrary{groupplots}
\pgfplotsset{compat=newest}
\usepackage{float}
\usepackage{footnote}
\usepackage{afterpage}

\begin{document}

\preprint{APS/123-QED}

\title{Optimizing Superconducting Qubit Performance: A Theoretical Framework for Design, Analysis, and Calibration}

\author{Sirshi S Ram}
\email{sirshi007@gmail.com}
\affiliation{Department of Physics, Sri Sathya Sai Institute of Higher Learning, Vidyagiri, Prasanthi Nilayam, Puttaparthi, Andhra Pradesh 515134, India}

\author{Muralikrishna Molli}
\email{muralikrishnamolli@sssihl.edu.in}
\affiliation{Department of Physics, Sri Sathya Sai Institute of Higher Learning, Vidyagiri, Prasanthi Nilayam, Puttaparthi, Andhra Pradesh 515134, India}

\author{Vamshi Mohan Katukuri}
\email{vamshi.katukuri@iao.fraunhofer.de}
\affiliation{Fraunhofer IAO, Nobelstraße 12, 70569 Stuttgart, Germany}

\author{Bharadwaj Chowdary Mummaneni}
\email{bharadwaj.chowdary.mummaneni@iao.fraunhofer.de}
\affiliation{Fraunhofer IAO, Nobelstraße 12, 70569 Stuttgart, Germany}
\date{\today}

\begin{abstract}
	Designing a qubit architecture is one of the most critical challenges in achieving scalable and fault-tolerant quantum computing as the performance of a quantum computer is heavily dependent on the coherence times, connectivity and low noise environments. Superconducting qubits have emerged as a frontrunner among many competing technologies, primarily because of their speed of operations, relatively well-developed and offer a promising path toward scalability. Here, we address the challenges of optimizing superconducting qubit hardware through the development of a comprehensive theoretical framework that spans the entire process - from design to the calibration of hardware through quantum gate execution. We develop this framework in four key steps: circuit design, electromagnetic analysis, spectral analysis, and pulse sequencing with calibration. We	first refine the qubit’s core components - such as capacitance, Josephson junctions, and resonators - to set the foundation for strong performance. The electromagnetic analysis, using the Lumped Oscillator model, allows us to map out the capacitance matrix, ensuring that we minimize spectral dispersion while maximizing coherence times. Following this, we conduct spectral analysis to fine-tune the qubit’s frequency spectrum and coherence properties, ensuring that the qubit parameters are optimized. Finally, we focus on pulse sequencing, including pulse-width optimization, DRAG optimization, and randomized benchmarking, to achieve high gate fidelity. We applied this framework to both Transmon and Fluxonium qubits, obtaining results that closely match those found in experimental studies. This work provides a detailed and practical approach to the design, optimization, and calibration of superconducting qubits, contributing to the broader effort to develop scalable quantum computing technologies.
\end{abstract}
\thispagestyle{empty}

\maketitle

\vspace{0.35cm}

\section{I. Introduction}
\label{sec:introduction}
Quantum computers have the potential to revolutionize the way we approach certain types of computational problems, particularly those that are intractable for classical computers. These include complex optimization problems, large-scale simulations of quantum systems, and certain cryptographic challenges \cite { Biamonte2017,chemistry,IEEE}. At the heart of this quantum revolution are superconducting qubits. Among the many competing technologies like trapped ions and cold atoms \cite{ions,JAKSCH200552} superconducting qubits have emerged as a frontrunner, primarily because of their speed of operations, relatively well-developed and offer a promising path toward scalability. Superconducting qubits leverage the properties of superconductors, materials that can conduct electricity without resistance at very low temperatures \cite{Lundy1989}. These qubits are typically based on Josephson junctions, which exploit the quantum tunneling of Cooper pairs (pairs of electrons bound together at low temperatures) to create discrete energy levels that can represent qubits \cite{Krantz2019}. 
  
Designing an optimal superconducting qubit architecture is one of the most critical challenges in achieving scalable and fault-tolerant quantum computing \cite{Krantz2019}. The performance of a quantum computer is heavily dependent on the coherence times, connectivity and low noise environments \cite{Fu2021}. All of these factors are influenced by a complex interplay of material properties, device geometry, and fabrication techniques \cite{5}. The search for the optimal superconducting qubit design began in earnest in 1999 \cite{Nakamura1999}, with early efforts focused on the Cooper-pair box qubit, which was later refined to create more robust qubits such as the Transmon, and Fluxonium \cite{Gao2021}. Each of these qubits represents an evolution in superconducting qubit technology, with improvements aimed at enhancing coherence times and reducing sensitivity to noise \cite{Krantz2019}. The Transmon qubit, for example, is a variation of the Cooper-pair box that offers reduced charge noise sensitivity, making it one of the most widely used qubits in today’s quantum computing ecosystem.
  
Despite these advancements, the challenge of realizing a practical, large-scale quantum computer remains formidable \cite{7927104}. The design of superconducting qubits for large-scale quantum computing is inherently an iterative process, requiring continuous optimization and calibration \cite{kunasaikaran2024frameworkdesignrealizationalternative}. This involves fine-tuning material properties, qubit dimensions, and chip design to achieve the right balance between strong coupling for control and readout and minimizing environmental noise.

\begin{figure*}[ht]
	\centering
	\includegraphics[width=1\linewidth]{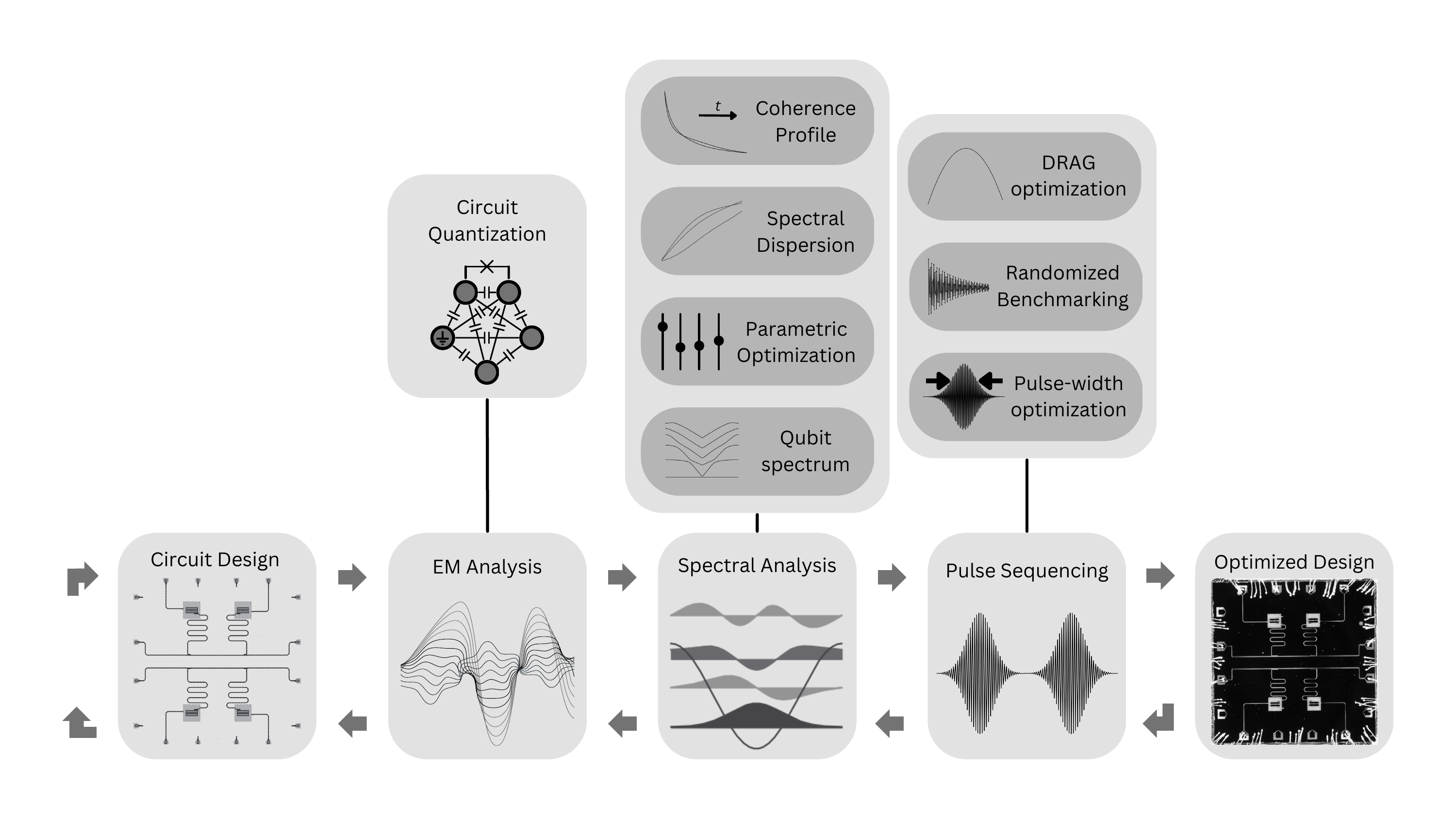}
	\caption{\justifying A pipeline outlining the procedure from qubit circuit design to pulse sequencing, involving an iterative process to converge on optimized processor parameters. (Image used in "Optimized Design" credits: Yuan et. al. \cite{28})
	}
	\label{fig:flowchart}
\end{figure*}
  
In this work, we aim to address the challenges of optimizing superconducting qubit hardware through the development of a comprehensive theoretical framework that spans the entire process—from design to the calibration of hardware through quantum gate execution. This approach aligns with established practices within the superconducting hardware community. The frame work consists of the following four steps, also see Figure \ref{fig:flowchart}:

\begin{enumerate}
	\item Circuit Design: The initial step focuses on designing the qubit circuit, optimizing key parameters like capacitance and Josephson junction inductance to minimize spectral dispersions and maximize coherence times.
	\item Electromagnetic Analysis: The second step involves a quasi-static analysis to extract the effective capacitance of the qubit circuit, which is crucial for circuit quantization. This analysis yields key qubit parameters, including frequency, anharmonicity, and cross-Kerr dispersions with neighboring systems. The Lumped Oscillator Model (LOM), detailed later, is applied for this purpose.
	\item Spectral Analysis: In the third step, spectral analysis is conducted to extract the qubit spectrum, coherence profile, and spectral dispersion. This step allows for the fine-tuning of qubit parameters to optimize performance.
	\item Pulse Sequencing and Calibration: The final step involves pulse sequencing, including optimization of pulse duration, DRAG (Derivative Removal by Adiabatic Gate) optimization, and randomized benchmarking. These techniques are used to calibrate the quantum gates and achieve maximum gate fidelity.
\end{enumerate}

We applied this framework to a single Transmon qubit and a single Fluxonium qubit, achieving results that closely align with practical outcomes reported in the literature \cite{ibm_quantum, PhysRevLett.130.267001}. 

The following section delves into the theoretical underpinnings of our research. We begin by discussing qubit circuit design, outlining the principles and techniques involved in constructing quantum circuits. To analyze the electromagnetic behavior of these circuits, we employ the Lumped Oscillator Model (LOM). Subsequently, we explore qubit control, a crucial aspect of quantum computing that involves pulse sequencing. Pulse sequencing is essential for manipulating qubit states and executing quantum gates. To ensure optimal performance, we meticulously benchmark qubit characteristics and optimize pulse sequences. The results of our theoretical investigations are presented in the subsequent section.

\section{II. Theory}

\subsection{A. The Qubit Circuit}
\label{sec:TMONandFLUX}

\begin{figure}[ht]
	\begin{subfigure}[b]{0.425\linewidth}
		\centering
		\includegraphics[width=1\linewidth]{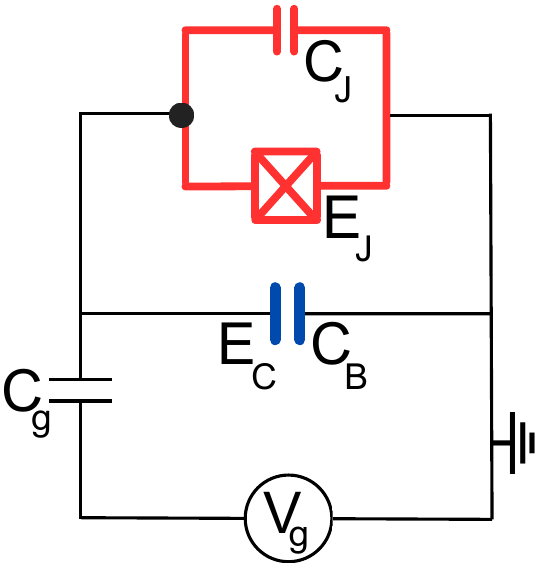}
		\caption{}
		\label{TMON_circuit1}
	\end{subfigure}  
	\hfill
	\begin{subfigure}[b]{0.465\linewidth}
		\centering
		\includegraphics[width=1\linewidth]{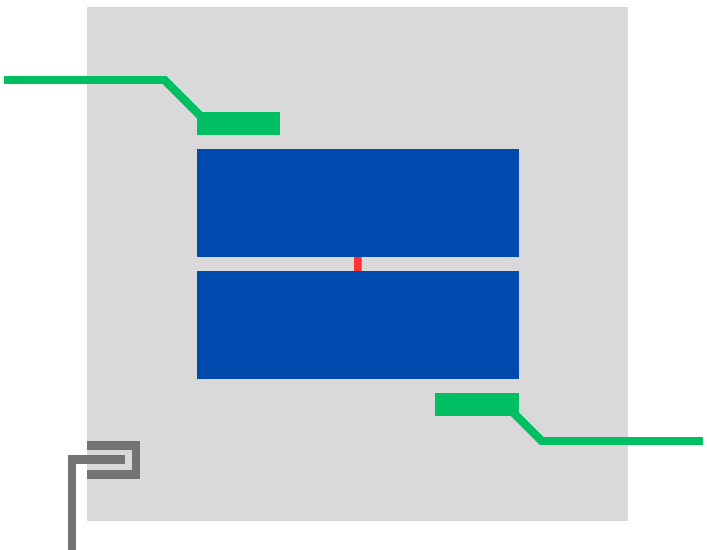}
		\caption{}
		\label{TMON_circuit2}
	\end{subfigure}   
	\caption{\justifying (a). Lumped element model of the Transmon qubit, featuring the Josephson junction (red metal strip, representing overlapping strip of aluminium islands from each island across the oxide layer, \cite{minev2021energyparticipationquantizationjosephsoncircuits}), and shunt capacitor (dark blue), all controlled by an electrostatic gate of voltage $V_g$ and capacitance $C_g$. (b). Transmon circuit layout, showing capacitive islands ('Top' and 'Bottom' pads, dark blue), Josephson junction (red), bus pads ('Readout' and 'Coupler' pads, green), and charge line (dark grey).}
	\label{TMON_circuit}
\end{figure}

The Transmon circuit \cite{16, 17} employs a nonlinear Josephson junction in its circuit design, see Figure \ref{TMON_circuit}. This is responsible for the cosine nature of the potential landscape, see Figure \ref{TMON_fig_wavefunc}. The Hamiltonian that describes a Transmon circuit in the charge basis $\{|n\rangle\}$, where $n$ is the number of Cooper-pairs of electrons, has the following structure: 
\begin{equation} \tag{1}
	\begin{aligned}
		\hat{H} &= \hat{H}_{\text{charge}} + \hat{H}_{\text{tunnel}} = \frac{\hat{Q}^2}{2C} - E_J\cos{(\hat{\Phi})} \\
		&= 4E_C(\hat{N} - n_g)^2 - E_J \cos(\hat{\Phi}),
	\end{aligned}
\end{equation}
where, the first (kinetic) part of Equation (1) corresponds to the charging process and the second (potential) part is the tunneling. Here, $E_C = e^2/2C$ is the charging energy, $C$ is the qubit capacitance, $\hat{N}$ is the charge number operator, $n_g$ is the number of charges at the external dc gate, see Figure \ref{TMON_circuit} (a), and $E_J$ is the Josephson energy \cite{Krantz2019, 16}.

The Transmon is a modified Cooper-Pair Box (CPB)~\cite{16}, which simply consists of a Josephson junction. The CPB employs an $E_J/E_C$ ratio of 1, resulting in periodic spectral crossings, imposing strict requirements on maintaining \(n_g\). On the other hand, Transmons employ ratios greater than 10 ~\cite{17}, which is achieved by the introduction of a large shunt capacitor, as shown in Figure ~\ref{TMON_circuit} (a). Such a qubit exhibits exponentially suppressed spectral dispersion, alleviating the need to precisely maintain $n_g$ and rendering it robust against fluctuating charge noise~\cite{31}. However, this improvement comes at the expense of reduced anharmonicity, which decreases polynomially with increasing $E_J/E_C$.

\begin{figure}[ht]
	\begin{subfigure}[b]{0.505\linewidth}
		\centering
		\includegraphics[width=1\linewidth]{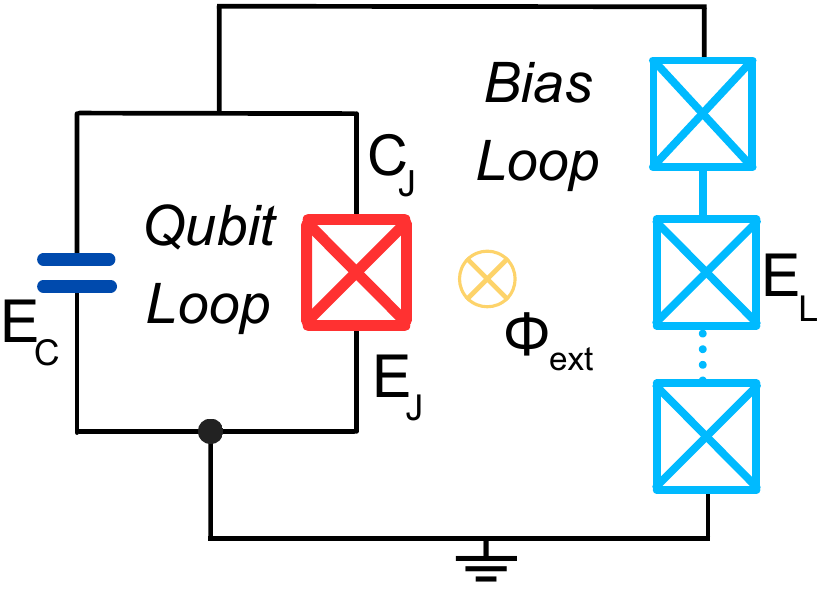}
		\caption{}
		\label{FLUX_circuit1}
	\end{subfigure}  
	\hfill
	\begin{subfigure}[b]{0.465\linewidth}
		\centering
		\includegraphics[width=1\linewidth]{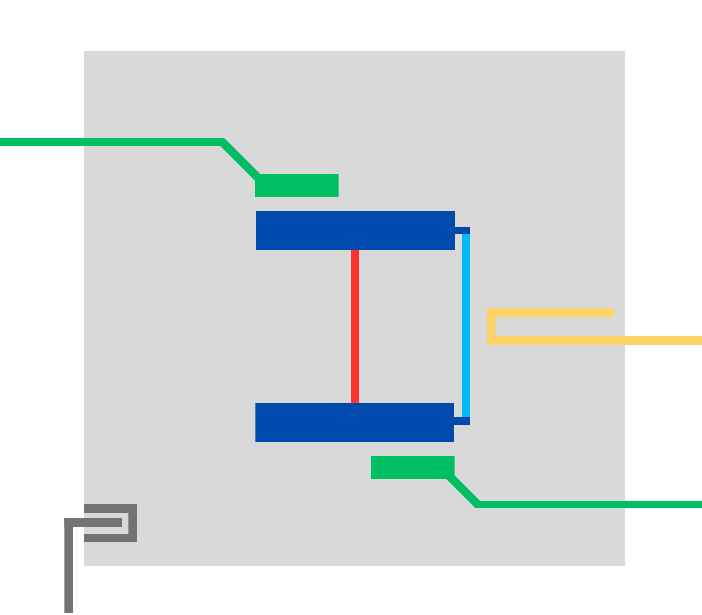}
		\caption{}
		\label{FLUX_circuit2}
	\end{subfigure}   
	\caption{\justifying (a). The Fluxonium qubit's lumped element model includes the Josephson junction (red), shunt capacitor (dark blue), forming the qubit loop, and the kinetic superinductor (light blue), all controlled by a magnetic bias $\phi_{ext} (= \Phi_{ext}/\Phi_0)$, applied across the bias loop. (b). The circuit layout of the fluxonium qubit displays the capacitive islands ('Top' and 'Bottom' pads, dark blue), Josephson junction (red), kinetic superinductor (light blue), magnetic bias line (yellow), bus pads ('Readout' and 'Coupler', green), and charge line (dark grey).}
	\label{FLUX_circuit}
\end{figure}

The Fluxonium qubit incorporates a large shunt kinetic inductor, as shown in Figure \ref{FLUX_circuit}(a), replacing the Transmon's shunt capacitor \cite{16, 18}. This component is typically fabricated by densely arranging hundreds of Josephson junctions \cite{16}. 
Its inclusion creates two loops: the qubit loop, which contains the Josephson element along with its geometric capacitance, and the bias loop, which includes the superinductor, as illustrated in Fig. \ref{FLUX_circuit}(a). The Hamiltonian for the Fluxonium is given by \cite{19}: \[\hat{H} = 4E_C\hat{N}^2 + \frac{1}{2}E_L \hat{\Phi}^2 - E_J \cos(\hat{\Phi} - \frac{\Phi_{\text{ext}}}{\Phi_0}). \tag{2}\] Here, $\hat{\Phi}$ is the flux operator, $\Phi_{\text{ext}}$ is the external magnetic flux applied across the bias loop of the Fluxonium \cite{15}, and $E_L = \Phi_0^2/L$ is the kinetic inductor energy. $E_L$ is responsible for the Fluxonium's enhanced resilience to both flux and charge noise. It acts as a short at low frequencies by dissipating noisy charges away from the qubit loop, thereby allowing $E_C$ to be tuned without compromising charge noise immunity \cite{18}. The high inductance $L$ opposes changes in current that are induced by noisy magnetic fields, thus reducing the qubit's sensitivity to bias-line noise. Typical energy ratios $E_J/E_C, E_J/E_L$ for the Fluxonium are within 2 and 10.\cite{21} The Fluxonium operates at the flux-frustration point, corresponding to a half-integer flux quantum $(\Phi_{\text{ext}}/\Phi_0 = 0.5)$ where $\Phi_0 = h/2e$ is the magnetic flux quantum \cite{16, 21}. Operating at the flux-frustration point results in a symmetric double-well potential in the Fluxonium system \cite{16}, see also Figure \ref{FLUX_fig_wavefunc}. In the heavy-Fluxonium regime (energy ratios $\sim 10$) exhibit qubit transition frequencies $\omega_{0\rightarrow1}$ on the order of tens of MHz and anharmonicities $\alpha$ on the order of GHz \cite{FastFlux}. These low-frequency transitions, known as fluxon transistions are a distinctive feature of Fluxonium qubits and arise due to the large inductive shunt, which enhances $\alpha$, and coherence times \cite{Hassani_2023, strickland}, coherence times are discussed in Section \hyperref[sec:coherencetimes] The enhanced $\alpha$ minimizes population leakage during microwave pulse driving \cite{18}. Transitions such as $\omega_{0\rightarrow2}$ and $\omega_{1\rightarrow3}$ are known as plasmon transitions \cite{PhysRevX.14.041014}. 
However, such low transition frequencies make the system more vulnerable to thermal noise \cite{16}. To mitigate this issue, light-Fluxoniums are employed (energy ratios $\sim 2$). These qubits are characterized by both $\omega_{0\rightarrow1}$ and $\alpha$ in GHz-scale. However, coherence times of the qubits in the heavy-Fluxonium regime are an order of magnitude larger than in qubits in the light-Fluxoniums regime \cite{PRXQuantum.4.040342}.

\subsection{B. Lumped Oscillator Model Analysis}
\label{sec:LOM}

The lumped oscillator model (LOM) analysis \cite{27} offers a numerical approach for extracting Hamiltonians of topologically complex quantum circuits. This approach employs the quasi-lumped approximation \cite{lumped, 27}, a hybrid technique that integrates the lumped modeling of qubit circuits with the distributed modeling of transmission lines. The lumped model relies on the quasi-static approximation \cite{novotny2012principles}, which is valid when the physical dimensions of the system are much smaller than the wavelength of its oscillations. This approximation allows the spatial variation of the driving fields within the structure to be neglected, simplifying the analysis of resonant systems by treating them as an effective LCR circuit \cite{16, pozar2011microwave}. Transmons and Fluxoniums, being micron-scale structures, are resonantly driven at RF-microwave wavelengths, ranging from millimeters to meters \cite{banwell1972fundamentals}, thus justifying this approach.

A quantum circuit with a defined geometry is partitioned into disjoint cells, each characterized by extractable electrostatic parameters, forming a quantum object within the topological Hilbert space \cite{27}. The electrostatic parameters are determined by solving Poisson's equation \cite{griffiths2017introduction}: \[ \grad^2 V(\vec{r}) = -\frac{\rho(\vec{r})}{\epsilon_0}, \tag{3}\] where \(\rho(\vec{r})\) represents the charge distribution of the structure. Solutions to Equation (3) yield the field intensities and potential distributions across the circuit, which can then be used to construct Maxwell's capacitance matrix \cite{dilorenzo2020maxwell}.

The general Lagrangian prescription for a superconducting qubit cell is given by: \[\mathcal{L}(\Phi_n, \dot{\Phi}_n) = \frac{1}{2} \dot{\Phi}_n^T C_n \dot{\Phi}_n - \frac{1}{2} \Phi_n^T L_n^{-1} \Phi_n + \sum_{j=1}^{J} \mathcal{E}_j(\Phi_j), \tag{4} \] where \( \Phi_n \) is the flux across the \(n^{\text{th}}\) node and serves as the generalized coordinate in the configuration space. Here, \( C_n \) and \( L_n \) are the Maxwell capacitance matrix and the linear $1/L$ inductance matrix of the \(n^{\text{th}}\) node, respectively. \(\mathcal{E}_j(\Phi_j) \) is the energy function of the \(j^{\text{th}}\) Josephson junction, which typically has a cosine potential.

The non-linear Josephson element functions as a tunneling junction, allowing Cooper pairs to traverse between capacitive islands of the qubit circuit. This process is governed by the flux-current relationship \cite{16} \[I(t) = I_c \sin\biggl({\frac{\Phi(t)}{\Phi_0}}\biggr), \tag{5}\] where $\Phi(t)$ is the flux across the junction, $\Phi_0$ is the flux quantum, and $I_c$ is the critical current of the Josephson junction. The energy function of the Josephson element is given by \cite{16, minev2021energyparticipationquantizationjosephsoncircuits}
\begin{equation} \tag{6}
	\begin{aligned}
		\mathcal{E}_J(\Phi) &= \frac{L_J I_c^2}{2}\biggl[1 - \cos^2\biggl(\frac{\Phi(t)}{\Phi_0}\biggr)\biggr] \\
		&= \frac{E_J}{2} \biggl[ \frac{1}{2} \biggl(\frac{\Phi}{\Phi_0}\biggr)^2 - \frac{1}{4!} \biggl(\frac{\Phi}{\Phi_0}\biggr)^4 + \mathcal{O}(\Phi^6) \biggr]
	\end{aligned}
\end{equation}
where $E_J = \Phi_0^2/L_J = I_c\Phi_0$ denotes the Josephson energy. The nodal flux is chosen as the generalized coordinate, with nodal charge as its conjugate momentum \cite{Nicholas, 16}, selected to prevent a multi-valued function when inverting Equation (5) \cite{minev2021energyparticipationquantizationjosephsoncircuits}. Capacitive elements correspond to kinetic energy, while inductive elements correspond to potential energy \cite{16}. Nodes formed between capacitive and inductive elements contribute to system dynamics; those between capacitor-capacitor or inductor-inductor do not. 

$C_n$ is extracted using a electrostatic finite-element method (FEM) solver, typically yielding a full-rank matrix. Such a solver approximates solutions to partial differential equations with defined geometry and boundary conditions. Cells are meshed into tetrahedron and hexahedron mesh-nodes and passed to a Poisson solver \cite{Schneider_2022} to compute $C_n$. The $1/L$ matrix is rank-deficient for Transmons and Fluxoniums, as the circuit houses only one linear inductor node, coming from first-order approximation of the Josephson potential. The rank-deficient nature of $1/L$ results in making the Lagrangian singular \cite{27}. LOM analysis addresses this by selectively removing non-dynamical nodes but retains relevant nodes (such as resonator nodes), verifying the reduced circuit faithfully maps to the original circuit.

The quantum Hamiltonian of the composite network is derived by applying a Legendre transformation to Equation (4) in a reduced basis, retaining only the essential nodes after selective removal. This Hamiltonian is then diagonalized to extract key qubit parameters, such as qubit transition frequency, anharmonicity, and cross-Kerr dispersion with other subsystems in the composite network. While advanced techniques like energy participation ratio (EPR) analysis offer a full-wave solution by solving Maxwell's equations from first principles, LOM analysis remains computationally efficient and sufficiently accurate for small-scale circuits, such as coupled-qubit circuits \cite{28, Squadds}.
\subsection{C. Coherence Times}
\label{sec:coherencetimes}

In experimental setups, qubits are vulnerable to stochastic environmental interactions, leading to noise and decoherence of the system \cite{16}. The interaction between the qubit and the noisy environment is expressed as $\hat{H}_\text{int} = v\hat{O}_q.\hat{\lambda}$, where $v$ indicates the interaction strength between the qubit and the environment, $\hat{O}_q$ is the qubit Hamiltonian, and $\hat{\lambda}$ represents the noisy environment of the qubit \cite{Krantz2019}. Typically, qubits are characterized by their coherence times $T_1$ and $T_2$, which are governed by the dominant noise channels \cite{16, 9}.

The longitudinal relaxation time $T_1$ describes the de-population of the excited level of the qubit \cite{16}. $T_1$ time is estimated using the Fermi's golden rule, which reads \[\Gamma_{1\rightarrow 0} = \frac{1}{T_1} = \frac{1}{\hbar^{2}} |\langle 0|\hat{S}| 1\rangle|^2 S_{\Lambda}(\omega_{1\rightarrow 0}), \tag{7}\] where $\hat{S}$ is the qubit-environment coupler of coupling type $\sigma_x$ or $\sigma_y$, $S_{\Lambda}(\omega_{1\rightarrow 0})$ is the noise power spectral density which characterizes the frequency distribution of the noise power for the $\Lambda^{th}$ noise channel\cite{16, 19}. The power spectral density \(S_{\Lambda}(\omega_{0 \rightarrow 1})\) is given by \cite{Krantz2019} \[S_{\Lambda}(\omega_{0 \rightarrow 1}) = \int_{-\infty}^\infty \langle \hat{\Lambda}(t) \hat{\Lambda}(0) \rangle e^{-i\omega_{0 \rightarrow 1} t} \mathrm{d}t. \tag{8}\]  This follows the Wiener-Khintchine theorem, which relates the spectral density to the spectral decomposition of the auto-correlation of the $\Lambda^{th}$ noise channel.

The transverse relaxation time $T_2$ describes the loss of relative phase in superposition states \cite{16}. $T_2$ is caused by both $T_1$ and pure dephasing time $T_\phi$. These times are related by the following relation \cite{Krantz2019, 16, 15, 9}: \[\frac{1}{T_2} = \frac{1}{T_\phi} + \frac{1}{2T_1}.\] $T_\phi$ is estimated by calculating the power spectral density of the form \[S_\lambda(f) = \frac{A_\lambda^2}{|f|} \tag{9}\] for $1/f$ noise sources \footnote{1/f noise is a low-frequency classical noise, symmetric in frequency. This happens due to interaction of the microscopic degrees of freedom of the device with other quantum variables of the system \cite{Paladino_2014}.} \cite{17, PhysRevB.72.134519, Paladino_2014}. In this expression, $A_\lambda$ represents the noise strength of the $\lambda^\text{th}$ pure dephasing channel, and $f$ is the frequency of the $1/f$ noise spectrum bound by an infrared cut-off and an ultraviolet cut-off \cite{PhysRevB.72.134519, 9}. The dephasing rate due to a linearly-coupled $\lambda^\text{th}$ noise channel is calculated using the following expression: \[\Gamma_\phi = \frac{1}{T_\phi} = \sqrt{2} A_\lambda\frac{\partial\omega_{0\rightarrow1}}{\partial\lambda}\sqrt{|\ln{\omega_\text{low}t_\text{exp}}|} \tag{10}\] where $\omega_{0\rightarrow1}$ is the qubit transition frequency, $\omega_\text{low}$ is the low cut-off frequency \cite{PhysRevB.72.134519, 9, Bylander_2011}. Here, $t_\text{exp}$ is the duration of the Ramsey interferometry or spin-echo sequences, which are typical experiments used in estimating $\Gamma_\phi$ \cite{19, Krantz2019, 16}. 

The effective coherence times is a sum of coherence times due to individual noise channels of the qubit, i.e., \[\frac{1}{T_1^\text{eff}} = \sum_{i=1}^N \frac{1}{T_1^i} \tag{11}\] \[\frac{1}{T_2^\text{eff}} = \sum_{j=1}^M \frac{1}{T_\phi^j} + \frac{1}{2}\sum_{i=1}^N \frac{1}{T_1^i} \tag{12}\] where $M$ and $N$ are the total number of pure dephasing and longitudinal relaxation noise channels inducing the qubit decoherence \cite{9, 33}.

\subsection{D. Quantum Control}
\label{sec:GateDriving}

Quantum control is essential in implementing and demonstrating quantum computing, enabling precise manipulation of qubit states through unitary operations. The qubit state is mathematically represented as a vector in a two-dimensional Hilbert space, which can be conveniently visualized on the Bloch sphere. On this sphere, unitary operations correspond to rotations around its axes through \(\mathrm{SU}(2)\) gates \cite{Krantz2019}.

A general Hamiltonian for a two-level system under a sinusoidal drive is given by:\[ \hat{H} = \hbar \epsilon \hat{\sigma}_z + \hbar \Omega(t) \cos(\omega_{\text{d}} \tau) \hat{\sigma}_j, \quad j \in \{x, y\}, \tag{13} \] where $\epsilon = \frac{\omega_{0 \rightarrow 1}}{2}$, $\omega_{0 \rightarrow 1}$ is the transition frequency from the state $\lvert 0 \rangle$ to $\lvert 1 \rangle$, $\Omega (t)$ is the driving pulse whose amplitude is a function of time $t$, $\omega_{\text{d}}$ is the drive frequency, and $\tau$ is the duration of applied drive. In the rotating frame at $\omega_{\text{d}}$, the Hamiltonian becomes:\[ \hat{H} = \hbar \left( \epsilon - \frac{\omega_{\text{d}}}{2} \right) \hat{\sigma}_z + \frac{\hbar \Omega(t)}{2} \hat{\sigma}_j, \quad j \in \{x, y\}. \tag{14}\] The expectation value $\langle \sigma_z \rangle$ provides the state of the qubit, where $\langle \sigma_z \rangle = \pm1$ indicates the qubit is in the state $\lvert 1 \rangle$ and state $\lvert 0 \rangle$ respectively. Under coherent driving, this expectation value exhibits a sinusoidal behavior, which can be expressed as: \[\langle \sigma_z \rangle = -\cos\bigl[\Omega(t) \tau\bigr]. \tag{15}\] 

Applying a resonant drive to a qubit state $|0\rangle$, the qubit undergoes a complete transition from the state $\lvert 0 \rangle$ to $\lvert 1 \rangle$ when $\Omega(t) \tau = \pi$, where $\tau $ is the pulse duration. This is the realization of a Pauli-X gate. The oscillations in Equation (15) are known as Rabi oscillations \cite{Dudin_2012}.

A general representation of an $\mathrm{SU}(2)$ gate is given by the universal gate:
\begin{equation} \tag{16}
	\begin{aligned}
		U(\theta, \phi, \lambda) &= R_z(\phi) R_y(\theta) R_z(\lambda)\\
		&=\begin{pmatrix}
			\cos(\theta/2) & -e^{i \lambda} \sin(\theta/2) \\
			e^{i \phi} \sin(\theta/2) & e^{i(\phi + \lambda)/2} \cos(\theta/2)
		\end{pmatrix},
	\end{aligned}
\end{equation}
(upto a global phase of $e^{-i\frac{\phi+\lambda}{2}}$) where,
$R_j(\theta) = e^{-i \frac{\theta}{2} \sigma_{j}}$ for $j \in \{X, Y, Z\}$ are the parameterized rotation gates about the Bloch sphere axes, with $\sigma_j$ representing the corresponding Pauli matrices.

Gate operations on superconducting qubits are realized through capacitively-driven nanosecond microwave pulses \cite{10.1063/5.0029735}. The longitudinal gates such as Pauli-X and Pauli-Y, are implemented by modulating the interaction between the states $\lvert 0 \rangle$ and $\lvert 1 \rangle$ at their frequency difference, $\omega_{0 \rightarrow 1} = (E_{\lvert 1 \rangle} - E_{\lvert 0 \rangle})/\hbar$ \cite{16}. The transverse gate, Pauli-Z, is realized by inducing a relative phase shift between the states $\lvert 0 \rangle$ and $\lvert 1 \rangle$. This phase shift can be achieved either by using detuned control fields or by applying composite longitudinal gates \cite{Krantz2019}.

Gate pulses are typically Gaussian-shaped, defined as \cite{McKay_2017}:
\begin{equation}  \tag{17}
	\Omega_{\text{Gaussian}}(t) = \Omega(t)\times\frac{e^{\frac{-t^2}{2\delta^2}} - e^{\frac{-\tau^2}{2\delta^2}}}{1 - e^{\frac{-\tau^2}{2\delta^2}}}, \quad \text{for} \; t \leq \tau,
\end{equation}
Here $\tau = 4\delta$, where $\delta$ denotes the standard deviation of the Gaussian \cite{McKay_2017}.

The performance of a quantum gate is quantified by its fidelity (\(\mathcal{F}\)), which measures the closeness between two quantum states \cite{jozsa1994fidelity} (i.e., probability intensity). For two density matrices, \(\hat{\rho}\) and \(\hat{\chi}\), the fidelity is defined as \cite{nielsen2010quantum}:
\[\mathcal{F} (\hat{\rho}, \hat{\chi}) = \left[\text{Tr}\left(\sqrt{\sqrt{\hat{\rho}} \hat{\chi}\sqrt{\hat{\rho}}}\right) \right]^2. \tag{18}\]
In the specific case where \(\hat{\chi} = |\psi\rangle \langle \psi|\) and \(\text{Tr}(\hat{\chi}^2) = 1\) (pure state) and \(\text{Tr}(\hat{\rho}^2) < 1\) (mixed state), the fidelity simplifies to $\mathcal{F} (\hat{\rho}, |\psi\rangle) = |\langle \psi | \hat{\rho} | \psi \rangle|^2.$

$\mathcal{F}$ plays a pivotal for assessing the accuracy of quantum gates, particularly in benchmarking and error analysis scenarios, as they provide a metric for evaluating the overlap between ideal and experimentally realized quantum states. 

High fidelity quantum gates are crucial for quantum information processing. To realize high fidelity gates, mitigation of errors arising from noisy qubits is necessary. In weakly anharmonic systems, the leakage of population to higher energy states contributes significantly to the qubit noise, lowering the gate fidelity. To mitigate such sources of errors, a DRAG term is introduced given by \cite{McKay_2017}:  
\begin{equation}  \tag{19}
	\Omega(t) = \Omega_{\text{Gaussian}}(t) + \beta \dot{\Omega}_{\text{Gaussian}}(t).
\end{equation}  
Here, $\beta$ is the DRAG coefficient. Optimizing \(\beta\) reduces leakage to higher energy levels, effectively enhancing pulse fidelity. This correction is demonstrated in \cite{hyyppä2024reducingleakagesinglequbitgates} in systems with limited anharmonicity.
\section{III. Results and discussion - bare qubit designs}
\label{sec:singlequbits}
This section presents our results in two parts, focusing on the design and performance of both Transmon and Fluxonium qubits. In Part \hyperref[parta]{A}, we examine the circuit designs of both qubits and analyze their spectral properties. In Part \hyperref[subsec:pulsing]{B}, analysis and gate-benchmarking of the qubits is presented.

\subsection{Part A: Design Analysis \& Optimization}
\label{parta}

Optimization of the qubit circuit parameters is crucial to enhance the performance of the qubit. We design, perform LOM analysis and optimize these parameters for the circuit layouts presented in Section \hyperref[sec:TMONandFLUX]{II.A.} using \texttt{Qiskit Metal (v0.1.5)} \cite{8, Minev2021_metal}. We use \texttt{scQubits} (v3.1.0) \cite{9} to evaluate the spectral properties and coherence characteristics for these qubits. A comprehensive optimization of physical design parameters, including pad dimensions, pad gaps, and substrate dimensions, is provided in Appendix-\hyperref[sec:BareQubits]{A}.

\subsubsection{A.1. Transmon}
The optimization routine for a Transmon qubit follows achieving a spectrally flat energy landscape for the two states of a qubit (qubit subspace), as a function of the bias gate charge ($n_g$), while maintaining sufficient anharmonicity ($\alpha$) and long effective relaxation times $T_1^\text{eff}$ and $T_2^\text{eff}$. This ensures minimized leakage into non-qubit subspace states and enhances the qubit's coherence budget ($\mathcal{C}$), defined as \[\mathcal{C} = \frac{\text{min}(T_1^\text{eff}, T_2^\text{eff})}{\tau}.\tag{20}\] Here, $T_1^\text{eff}$ and $T_2^\text{eff}$ denote the effective longitudinal and transverse relaxation times respectively, as discussed in Section \hyperref[sec:coherencetimes]{II.C.}.

\begin{figure}[ht]
	\centering
	\begin{subfigure}[b]{0.5\textwidth}
		\centering
		\includegraphics[width=\textwidth]{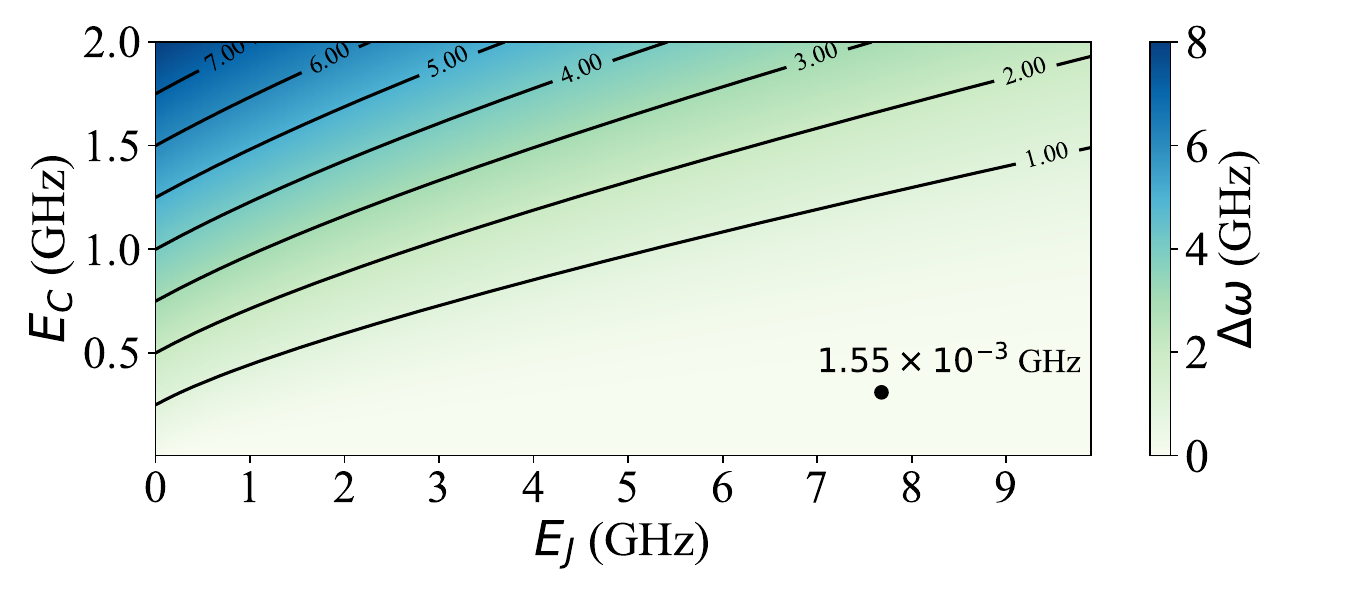}
		\caption{}
	\end{subfigure}
	\hfill
	\begin{subfigure}[b]{0.45\textwidth}
		\centering
		\input{Bare-qubit-analytics/bare_tmon}
	\end{subfigure}
	\caption{\justifying (a) The spectral dispersion $\Delta\omega$ when sweeping $E_J$ and $E_C$. The black dot indicates the minimized spectral dispersion of 1.55 MHz, for values of $E_J$ and $E_C$ at 7.68 GHz and 0.31 GHz respectively. (b). The flat-response of $\omega_{0\rightarrow1}$ to $n_g$ and (c) an exponential fall-off of $|0\rangle \rightarrow |1\rangle$ spectral dispersion.}
	\label{TMON_fig_specs}
\end{figure}

\begin{figure}[ht]
	\centering
	\input{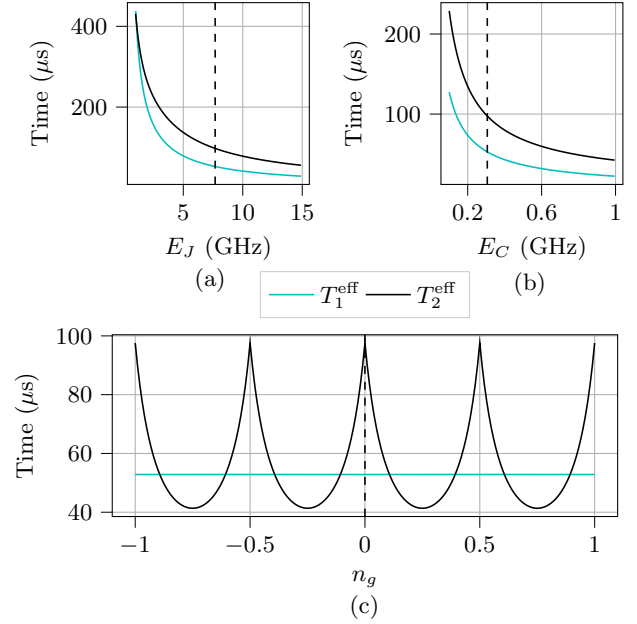}
	\caption{\justifying The effective coherence times of the Transmon is shown when sweeping (a) \(E_J\), (b) \(E_C\), and (c) $n_g$. The black dashed lines indicate the optimized \(E_J\) and \(E_C\) values, and the biasing point respectively.}
	\label{TMON_fig_coherences}
\end{figure}

To achieve qubit transitions independent of gate charge ($n_g$), one must examine the spectrum's periodicity at integer values of $n_g$. The transition frequency $\omega_{0\rightarrow1}$ reaches its maximum at $n_g = 0$ and its minimum at $n_g = 0.5$. By parametrically sweeping $E_J$ and $E_C$, the spectral fluctuations $\Delta\omega = \omega_{0\rightarrow1}^{n_g = 0} - \omega_{0\rightarrow1}^{n_g = 0.5}$ while keeping the qubit frequency within the desired range. As explained in Section \hyperref[sec:TMONandFLUX]{II. A.}, $\Delta\omega\rightarrow 0$ comes at the cost of reduced anharmonicity, see Figure \ref{TMON_fig_specs} (c) and Figure \ref{TMON_fig_wavefunc} (a). Conversely, lowering $E_J$ can enhance $T_1^\text{eff}$ (as seen in Figure \ref{TMON_fig_coherences}(a)) but risks approaching the CPB regime ($E_J/E_C \rightarrow 1$) \cite{16}. 

The Transmon device was modeled using the parameters listed in Table \ref{tab:table1}. Within this optimization approach, $\Delta\omega$ is minimized to $\sim 1.55$ MHz, as demonstrated in Figure \ref{TMON_fig_specs} (a), resulting in $\omega_{0\rightarrow1} = 4.00$ GHz and $\alpha = -397.10$ MHz. 

\begin{table}[ht]
	\centering
	\caption{Transmon circuit design}
	\label{tab:table1}
	\begin{tabular}{l|l}
		\toprule
		\textbf{Island dimension ($\mu m^2$)} & 400 $\times$ 125\\
		\textbf{Effective Josephson energy $E_J$} (GHz) & 7.68\\
		\textbf{Effective capacitive energy $E_C$} (GHz)& 0.31\\
		\textbf{Substrate dimension ($\mu m^2$)} & 600 $\times$ 600\\
		\textbf{Connector pad dimension ($\mu m^2$)} & 100 $\times$ 30\\
		\bottomrule
	\end{tabular}
\end{table}

Given that dilution refrigerators operate at 20 mK \cite{temp_20mK}, which corresponds to thermal excitations of 416.73 MHz, designing qubit frequencies in the 4–8 GHz range is crucial for mitigating thermal noise and maintaining coherence. 

Figure \ref{TMON_fig_specs} (c) demonstrates the invariance of  $\omega_{0\rightarrow1}$ with $n_g$. The qubit sensitivity to charge noise is illustrated in Figure \ref{TMON_fig_specs} (d), where the decay in spectral dispersion of the qubit follows an exponential decay for qubit subspace transitions. This behaviour implies a reduced sensitivity to higher $E_J/E_C$ ratios. 

The parameters $E_J$ and $E_C$ for the Transmon design are further refined by systematically measuring and optimizing its relaxation times. As seen in Figure \ref{TMON_fig_coherences} (a) and \ref{TMON_fig_coherences} (b), increasing \(E_J\) and \(E_C\) typically results in a decrease in both \(T_1^\text{eff}\) and \(T_2^\text{eff}\). This observation highlights the necessity of a nuanced optimization approach: higher values of \(E_J\) can reduce \(\Delta \omega\), but they may adversely affect \(T_1^\text{eff}\) and \(T_2^\text{eff}\). Conversely, reducing \(E_J\) can enhance coherence times, but with an increased \(\Delta \omega\). Thus, optimizing \(E_J\) involves balancing the reduction of \(\Delta \omega\) with the need to maximize \(T_1^\text{eff}\) and \(T_2^\text{eff}\) for sweeps of $E_J$ and $E_C$ respectively.

The effective decoherence times, \(T_1^\text{eff}\) and \(T_2^\text{eff}\), are summarized in Table \ref{tab:table2} and illustrated in Figure \ref{TMON_fig_coherences} (c). Here, \(T_1^\text{eff}\) is primarily dominated by the dielectric noise at oxide layer in the Josephson junction \cite{32, 33}. The critical current noise at the Josephson junction, and the fluctuating charge noise at the bias contributes to the qubit's pure dephasing \(T_\phi^\text{eff}\) \cite{33}. As seen in Figure \ref{TMON_fig_coherences} (c), the relationship \(T_2^\text{eff} \simeq 2T_1^\text{eff}\) holds at \(n_g = 0\), indicating \(T_\phi^\text{eff} \simeq 1.27\) ms. In contrast, at \(n_g = 0.25\), \(T_\phi^\text{eff}\) is reduced to approximately 68 \(\mu\)s. While the qubit exhibits insensitivity to relaxation caused by charge noise across different \(n_g\) values, it remains vulnerable to dephasing from charge noise. More details are provided in Appendix-\hyperref[sec:BareQubitsCoherence]{B}.  

\begin{table}[ht]
    \centering
    \caption{Transmon profile at $n_g = 0$}
    \label{tab:table2}
    \begin{tabular}{c|c|c|c|c}
    \toprule
    \textbf{$\omega_{0\rightarrow 1}$} (GHz) & \textbf{$\Delta\omega$} (MHz) & \textbf{$\alpha$} (MHz) & \textbf{$T_1^{\text{eff}}$} ($\mu$s) & \textbf{$T_2^{\text{eff}}$} ($\mu$s) \\
    \midrule
    \text{4.00} & \text{1.55} & \text{397.10} & \text{52.78} & \text{97.48} \\
    \bottomrule
    \end{tabular}
\end{table}

The optimized parameters of the Transmon, listed in Table \ref{tab:table2} is in good agreement with those reported for IBM's quantum processors \cite{ibm_quantum}.

\subsubsection{A.2. Fluxonium}

The optimization of the Fluxonium device focuses on fine-tuning three critical parameters: the Josephson energy (\(E_J\)), charging energy (\(E_C\)), and inductive energy (\(E_L\)). This is done particularly at the flux-frustration point (\(\Phi_{\text{ext}}/\Phi_0 = 0.5\)). The objective is to configure these parameters to model a Fluxonium qubit, characterized by $\omega_{0\rightarrow1}$ in the range of hundreds of MHz and $\alpha$ in the range of GHz, as discussed in Section \hyperref[sec:TMONandFLUX]{II. A}. 

\begin{table}[ht]
	\centering
	\caption{Fluxonium circuit design}
	\label{tab:table3}
	\begin{tabular}{l|l}
		\toprule
		\textbf{Island dimension ($\mu m^2$)} & 250 $\times$ 45\\
		\textbf{Effective Josephson energy $E_J$} (GHz) & 4.80 \\
		\textbf{Effective capacitive energy $E_C$} (GHz) & 0.99 \\
		\textbf{Effective inductive energy $E_L$} (GHz) & 0.89 \\
		\textbf{Substrate dimension ($\mu m^2$)} & 600 $\times$ 600\\
		\textbf{Connector pad dimension ($\mu m^2$)} & 50 $\times$ 30\\
		\bottomrule
	\end{tabular}
\end{table}

The Fluxonium model is defined by the design parameters listed in Table \ref{tab:table3}. \(E_J\) and \(E_C\) are optimized to 4.80 GHz and 0.99 GHz, respectively, while the inductive energy (\(E_L\)) is set to 0.89 GHz. This configuration results in a qubit transition frequency of \(\omega_{0 \rightarrow 1} = 305.17\) MHz and an anharmonicity of \(\alpha = 3.64\) GHz, as summarized in Table \ref{tab:table4}.

\begin{figure}[ht]
	\centering
	\input{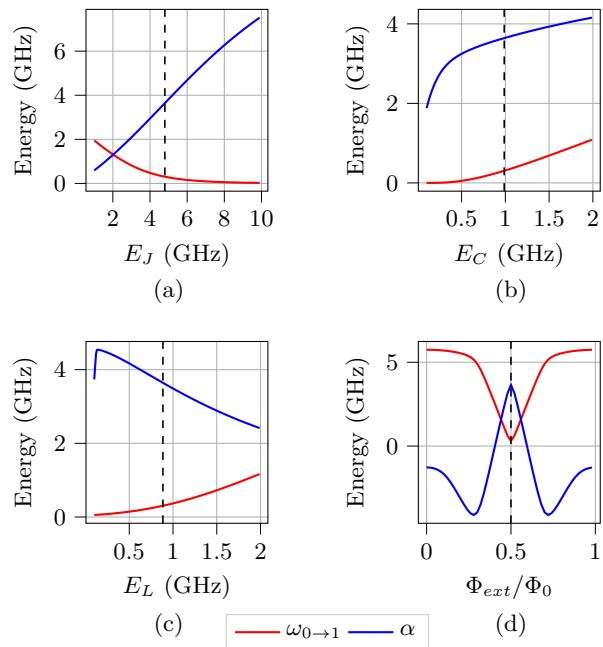}
	\caption{\justifying The transition frequency \(\omega_{0 \rightarrow 1}\) and anharmonicity $\alpha$ of the Fluxonium when (a) sweeping $E_J$ with $E_C$ and $E_L$ fixed to 0.99 GHz and 0.89 GHz respectively; (b) sweeping $E_C$ with $E_J$ and $E_L$ fixed to 4.80 GHz and 0.89 GHz respectively; (c) sweeping $E_L$ with $E_J$ and $E_C$ fixed to 4.80 GHz and 0.99 GHz respectively. (d) The qubit exhibits the desired low $\omega_{0 \rightarrow 1}$ and high $\alpha$ at the flux-frustration point.}
	\label{FLUX_fig_PERF}
\end{figure}

The resulting qubit frequency which we obtained corresponds to a temperature of 14.65 mK which might lead to a significant population in the non-qubit subspace for typical dilution fridge temperatures ($\sim$ 20 mK). However, this can be overcome by proper calibration of the gates \cite{19}, which is discussed in Section \hyperref[subsec:pulsing]{III. B}.

As shown in Figure~\ref{FLUX_fig_PERF} (a), increasing $E_J$, i.e., which deepens in the double-well potential, results in a reduction of the qubit transition frequency $\omega_{0\rightarrow 1}$ and an increase in the anharmonicity $\alpha$. Conversely, decreasing $E_L$ and $E_C$ also reduces $\omega_{0\rightarrow 1}$, see Figure \ref{FLUX_fig_PERF} (b) and \ref{FLUX_fig_PERF} (c). However, $\alpha$ shows opposite behavior with decreasing $E_C$ and $E_L$. Interestingly, $E_J/E_C \ge 6$, the qubit states become degenerate at the flux-frustration point ($\omega_{0\rightarrow 1} \rightarrow 0$). The values $E_J/E_C \sim 4.86$ and $E_J/E_L \sim 5.42$ are within the typical energy scales for Fluxoniums, as discussed in Section \hyperref[sec:TMONandFLUX]{II. A}. The behavior of $\omega_{0\rightarrow 1}$ and $\alpha$ with flux bias $\Phi_\text{ext}/\Phi_0$ is shown in Figure \ref{FLUX_fig_PERF} (d), where it can be seen that $\omega_{0\rightarrow 1}$ is lowest and $\alpha$ is highest about the flux-frustration point. 

\begin{figure}[ht]
	\centering
	\input{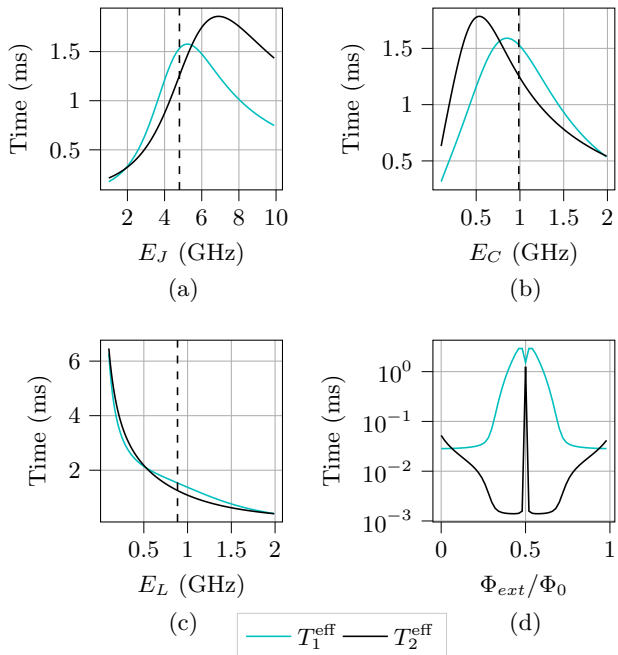}
	\caption{\justifying The effective coherence times of the Fluxonium at the flux-frustration point when (a) sweeping $E_J$ with $E_C$ and $E_L$ fixed to 0.99 GHz and 0.89 GHz respectively; (b) sweeping $E_C$ with $E_J$ and $E_L$ fixed to 4.80 GHz and 0.89 GHz respectively; and (c) sweeping $E_L$ with $E_J$ and $E_C$ fixed to 4.80 GHz and 0.99 GHz respectively. (d) The qubit exhibits maximum $T_1^\text{eff}$ around the flux-frustration point ($0.48 \Phi_0$ and $0.52 \Phi_0$), and $T_2^\text{eff}$ at the flux-frustration point.}
	\label{FLUX_fig_coherences}
\end{figure}

The effective coherence times calculated for the Fluxonium model are listed in Table \ref{tab:table4}. In Figure~\ref{FLUX_fig_coherences}, we show the behavior of $T_1^\text{eff}$ and $T_2^\text{eff}$ as a function of $E_J$, $E_C$, $E_L$ and $\Phi_\text{ext}/\Phi_0$. Details on the decoherence noise channels in Fluxonium model are provided in Appendix-\hyperref[sec:BareQubitsCoherence]{B.2}. $T_1^\text{eff}$ increases with increasing $E_J$, as illustrated in Figure \ref{FLUX_fig_coherences} (a), or with decreasing $E_C$ and $E_L$, as shown in Figure \ref{FLUX_fig_coherences} (b) and \ref{FLUX_fig_coherences} (c). However, lower values of \(E_C\) and \(E_L\) reduces qubit's sensitivity to charge noise and flux noise respectively, see Figure \ref{FLUX_fig_relaxations}.

As shown in Figure \ref{FLUX_fig_coherences} (d), a deviation of \(\Phi_\text{ext}/\Phi_0 = 0.3\) from the flux-frustration point nearly doubles \(T_1^\text{eff}\) but reduces \(T_2^\text{eff}\) by approximately three orders of magnitude. The lowest \(T_2^\text{eff}\) value is observed at a deviation of \(0.2 \Phi_0\). The coherence times of the Fluxonium are consistent with those reported in \cite{PhysRevLett.130.267001}.

\begin{table}[ht]
	\centering
	\caption{Fluxonium profile at $\Phi_{ext}/\Phi_0 = 0.5$}
	\label{tab:table4}
	\begin{tabular}{c|c|c|c}
		\toprule
		\textbf{$\omega_{0\rightarrow 1}$} (MHz) & \textbf{$\alpha$} (GHz) & \textbf{$T_1^{\text{eff}}$} (ms) & \textbf{$T_2^{\text{eff}}$} (ms) \\
		\midrule
		\text{305.17} & \text{3.64} & \text{1.53} & \text{1.26} \\
		\bottomrule
	\end{tabular}
\end{table}

\subsection{Part B: Pulse Sequencing and Benchmarking}
\label{subsec:pulsing}

As discussed in Section \hyperref[sec:GateDriving]{II.D}, quantum gates are implemented at the pulse level, operating on the nanosecond scale. Optimizing pulse parameters is essential for achieving effective quantum control. In this section, we present the pulse calibration routine and randomized benchmarking for two gates: the \(\pi\)-pulse and the Hadamard pulse. We utilized \texttt{seQuencing (v1.2.0)} \cite{sequencing} to define the pulses and \texttt{QuTiP (v4.7.1)} \cite{qutip} to solve the Lindblad master equation \cite{Brasil_2013} using the \textit{mesolve} function. This approach enabled the computation of non-unitary population dynamics for randomized and leakage benchmarking, as well as the evaluation of gate fidelities.

\begin{figure*}[ht]
	\centering
	\begin{subfigure}[b]{0.45\linewidth}
		\centering
		\includegraphics[width=1\linewidth]{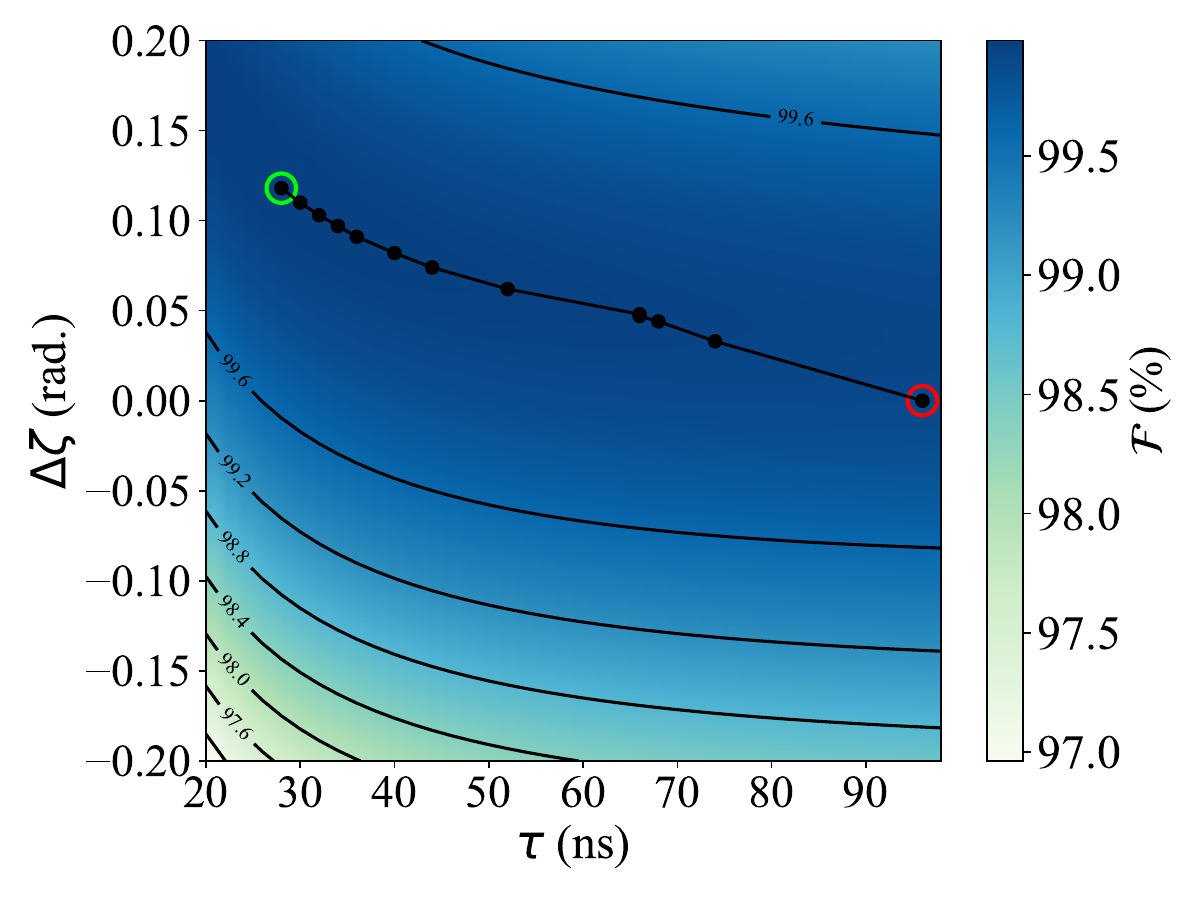}
		\caption{}
		\label{TMON_pi-calibration}
	\end{subfigure}  
	\hspace{0.4 cm}
	\begin{subfigure}[b]{0.45\linewidth}
		\centering
		\includegraphics[width=1\linewidth]{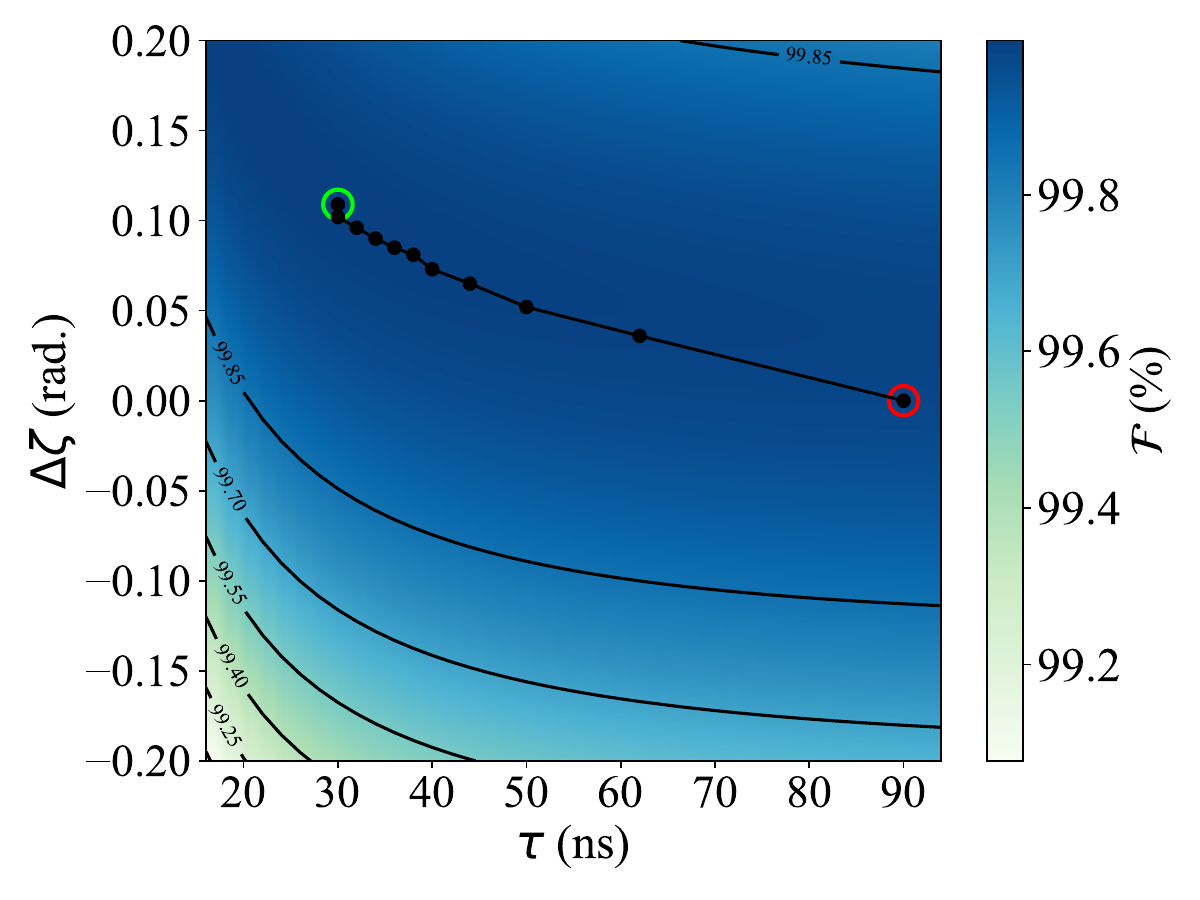}
		\caption{}
		\label{TMON_H-calibration}
	\end{subfigure}   
	\caption{\justifying A contour plot of $\Delta\zeta$ versus $\tau$ for the calibration of: (a) the $\pi$-pulse driven on the Transmon design. By slightly detuning from the ideal $\pi$ value, the Rabi oscillations are enhanced, reaching a maximum at $\Delta\zeta = +0.088$ radians. (b) The Hadamard pulse driven on the Transmon design, with the $\Delta \zeta$ optimized at 0.101 radians. The black line with triangular markers in both plots indicates the convergence towards the optimized $\tau$ and corresponding $\Delta\zeta$.}
	\label{TMON_pulse_cal}
\end{figure*}

\subsubsection{B.1. Transmon benchmarking}
\label{sec:TMON_benchmark}

When the Transmon model with a transition frequency $\omega_{0\rightarrow 1}$ of 4.00 GHz is initialized to state \(|0\rangle\) at a temperature of 20 mK, the state population is approximately 99.9930\% as dictated by the Boltzmann distribution \cite{lvov2024thermometry}. This implies that a thermal population ($p_\text{th}$) of approximately 0.007\% leaks into the higher energy states during initialization. For $T_1^\text{eff} = 52.78 \mu s$ (see Table \hyperref[tab:table2]{II}) The associated excitation rates and relaxation rates are \(\Gamma_\uparrow = p_\text{th}/T_1^\text{eff} = 1.89\) Hz, and \(\Gamma_\downarrow = (1 - p_\text{th})/T_1^\text{eff} = 18.95\) kHz respectively. The low \(\Gamma_\uparrow\) indicates minimal excitation to higher energy states, while the higher \(\Gamma_\downarrow\) signifies rapid relaxation to the ground state.

\begin{figure}[ht]
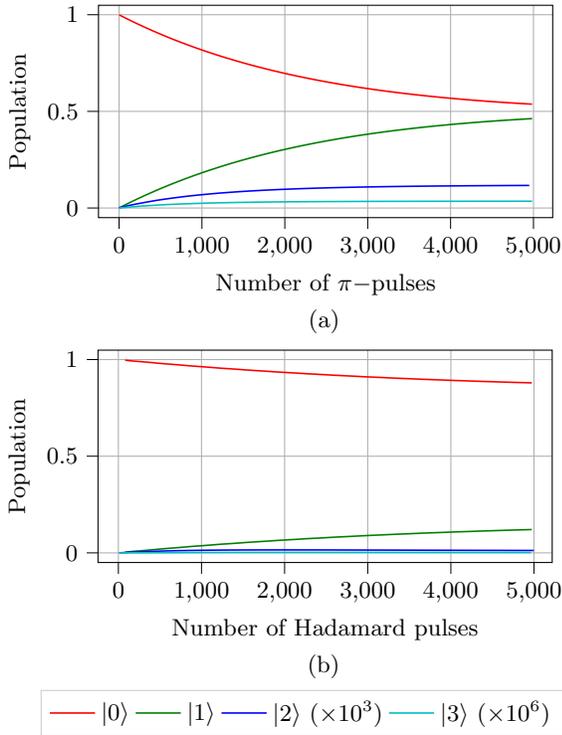

	\centering
	\begin{subfigure}[b]{0.475\textwidth}
		\centering
		\input{Bare-qubit-analytics/TMON_popdyn_X}
		\label{TMON_fig_popdyn_pi-pulse}
	\end{subfigure}
	\hfill
	\begin{subfigure}[b]{0.475\textwidth}
		\centering
		\input{Bare-qubit-analytics/TMON_popdyn_H}
		\label{TMON_fig_popdyn_H-pulse}
	\end{subfigure}
	\caption{\justifying Randomized benchmarking of the Transmon design under a (a). $\pi$-pulse sequence, and (b). a Hadamard gate sequencing. Population leakage into $|2\rangle$ is scaled by $10^3$ and $|3\rangle$ by $10^6$.}
	\label{TMON_popdyn}
\end{figure}

Following initialization, the gate calibration is performed by parametrically sweeping both the pulse duration (\(\tau\)) and the polar-rotational argument (\(\theta\)) to optimize gate fidelity. The calibration routine for the \(\pi\)-pulse \(U(\theta = \pi, \phi = 0, \lambda = \pi)\) (see Equation (16)) is executed as follows:

\begin{enumerate}
	\item $\tau$ is parametrically swept to identify the value that maximizes gate fidelity. 
	\item For the optimal \(\tau\), both \(\theta\) and \(\lambda\) are varied within the ranges \([\pi - \Delta\zeta, \pi + \Delta\zeta]\), where \(\Delta\zeta \ll \pi\). $\Delta\zeta$ introduces a slight detune to enhance the Rabi oscillations.
\end{enumerate}

These two steps are performed iteratively, allowing the process to converge on the optimal values of \(\Delta\zeta\) and \(\tau\). This iterative refinement enhances the fidelity \(\mathcal{F}\) (see Equation (18)) of the \(\pi\)-pulse and extends the coherence budget \(\mathcal{C}\) (see Equation (20)).

\begin{figure*}[ht]
	\begin{subfigure}[b]{0.45\linewidth}
		\centering
		\includegraphics[width=1\linewidth]{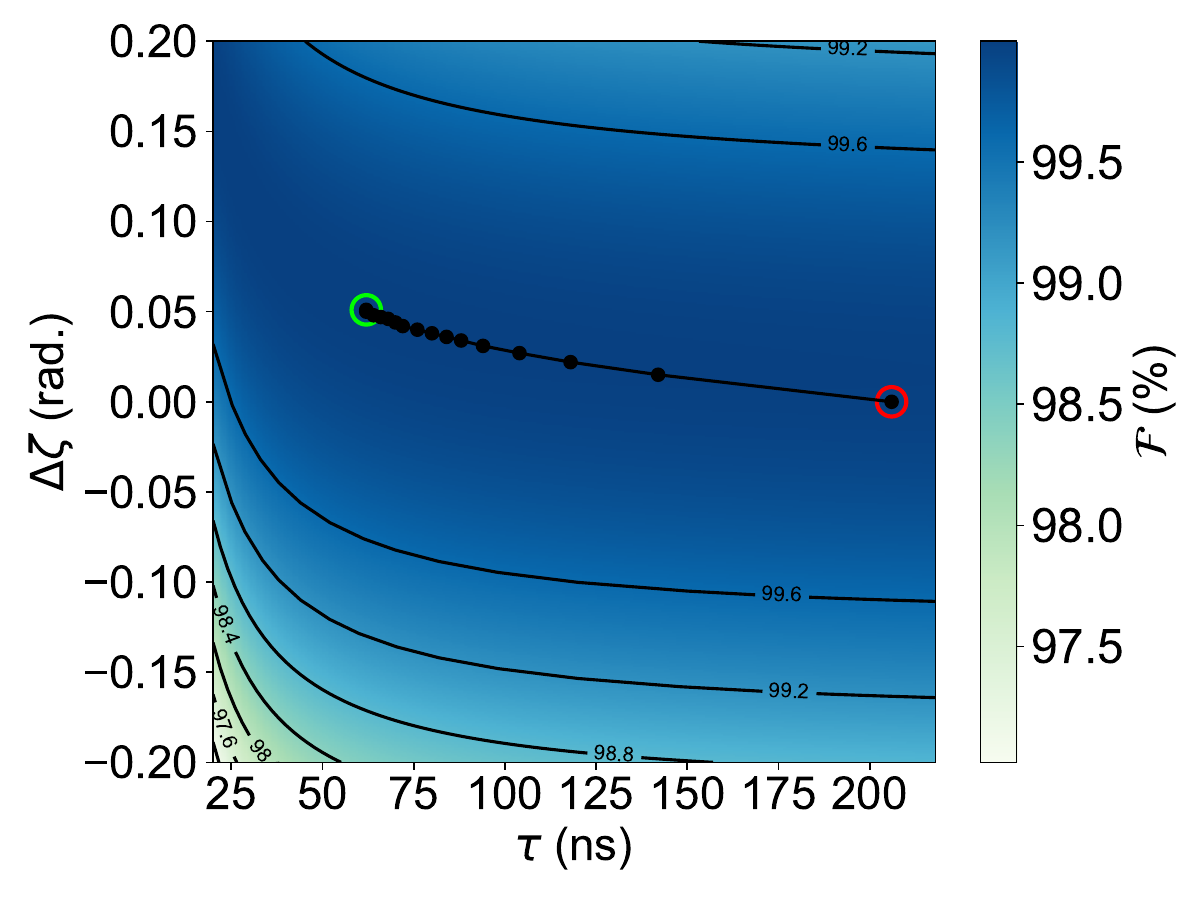}
		\caption{}
		\label{FLUX_pi-calibration}
	\end{subfigure}  
	\hspace{0.4cm}
	\begin{subfigure}[b]{0.45\linewidth}
		\centering
		\includegraphics[width=1\linewidth]{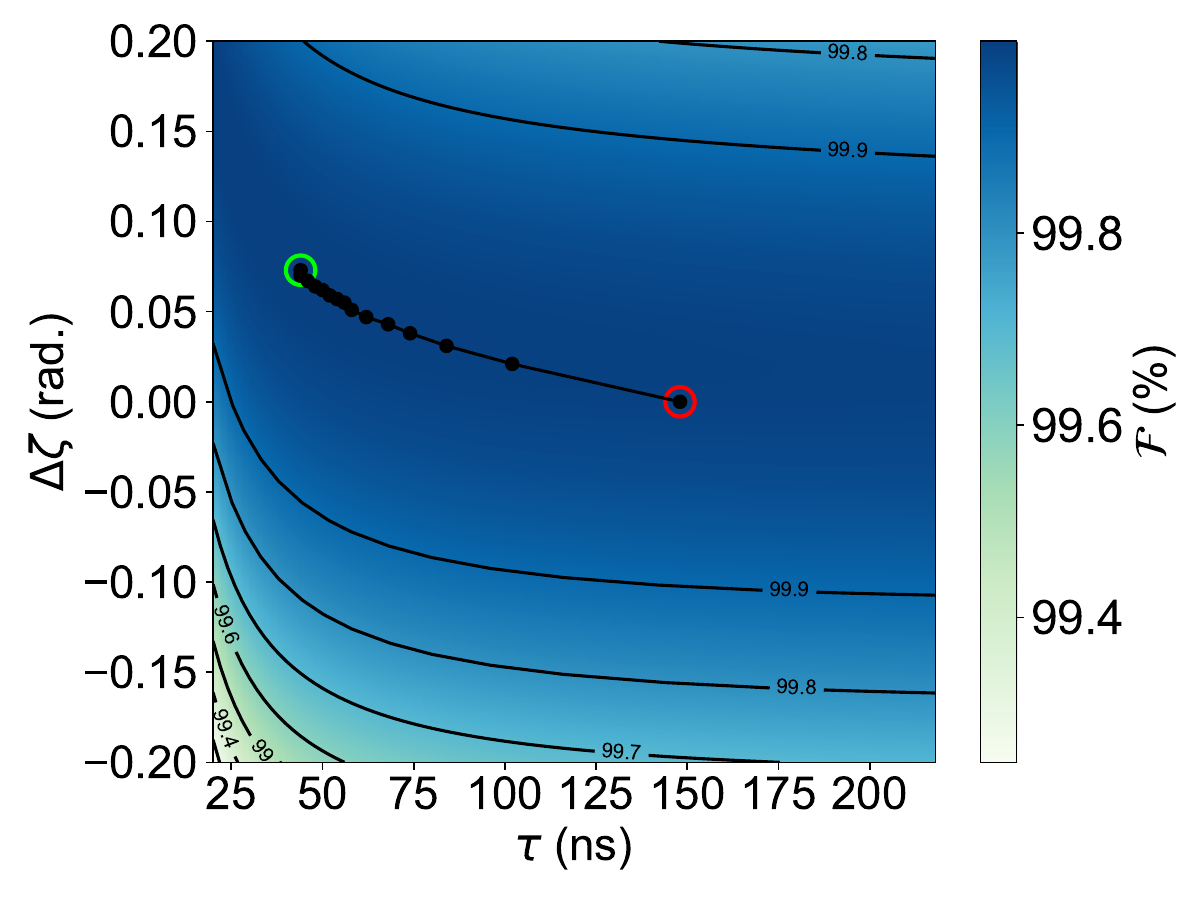}
		\caption{}
		\label{FLUX_H-calibration}
	\end{subfigure}   
	\caption{\justifying A contour plot of $\Delta\zeta$ versus pulse width for the calibration of (a). the $\pi$-pulse and (b) the Hadamard-pulse driven on the Fluxonium design. The black line in the plot illustrates the convergence towards the optimized $\tau$ and corresponding $\Delta\zeta$.}
	\label{FLUX_pulse_cal}
\end{figure*}

The calibration process is illustrated in Figure \ref{TMON_pulse_cal}(a), where the convergence path is marked by a black line with dotted markers. Initially, at \(\Delta\zeta = 0\), \(\tau\) for optimized $\mathcal{F}$ is 96.0 ns, achieving \(\mathcal{F} = 99.8840\%\) as highlighted by the red circle in the Figure \ref{TMON_pulse_cal}(a). Subsequent refinement reduces \(\tau\) to the optimal value of 28.0 ns (\(\delta = 7.0\) ns) at \(\Delta\zeta = +0.118\) radians, as indicated by the green circle. This optimization enhances \(\mathcal{F}\) to 99.9779\% and increases the corresponding \(\mathcal{C} = 1885\) gates, calculated using Equation (20). At the $1885^\text{th}$ pulse, the population of state \(|0\rangle\) decreases to 70.6\%.

The randomized benchmarking of the Transmon qubit involves applying a sequence of optimized pulses in rapid succession and analyzing the qubit's population dynamics to estimate the average error per gate (\(\epsilon\)) \cite{Knill_2008, Emerson_2005}. Figure \ref{TMON_popdyn} (a) presents the population dynamics of the Transmon qubit during the $\pi-$pulse benchmarking sequence. The state \(\lvert 0 \rangle\) shows a gradual accumulation of error, and simultaneously leaking into non-computational states \(\lvert 2 \rangle\) and \(\lvert 3 \rangle\) (indicated by the blue line and the cyan line respectively). This is due to the interaction of the qubit coupling with the environment, see Section \hyperref[sec:coherencetimes]{II. C}. \(\epsilon\) is estimated for the \(\pi\)-pulse by fitting an exponential function of the form $f(x) = A e^{xt} + y_0$ to the decay trend of state \(\lvert 0 \rangle\). Here, $A$ is the amplitude, $t$ is the time, and $y_0$ is the offset. \(\epsilon\) is calculated to be approximately \(4.3\times10^{-3}\).

The Hadamard gate, defined as \(H = U(\theta = \pi/2, \phi = 0, \lambda = \pi)\), facilitates access to the equatorial plane of the Bloch sphere. The calibration process begins with the qubit initialized in its ground state (\(\lvert 0\rangle\)), followed by the application of a Hadamard gate to create the superposition state \(\lvert +\rangle = (|0\rangle + |1\rangle)/\sqrt{2}\). Figure \ref{TMON_pulse_cal} (b) illustrates the calibration process for the Hadamard gate pulse. The black line with dotted markers indicates that the initial $\tau$ at \(\Delta\zeta = 0\) is 90.0 ns, with $\mathcal{F} = 99.9763\%$, indicated by the red circle. Through iterative refinement, the pulse duration is reduced to an optimal value of 30.0 ns (\(\delta = 7.5\) ns) at \(\Delta\zeta = +0.109\) radians, resulting in an improved $\mathcal{F} = 99.9950\%$, indicated by the green circle. The randomized benchmarking for the Hadamard gate is shown in Figure \ref{TMON_popdyn} (b). After 5000 gates, the ground state population decreases to \(87.93\%\). The coherence budget $\mathcal{C}$ is calculated to be 1759 gates and the ground state population after 1759 gates is approximately \(93.91\%\). The optimized pulse has an \(\epsilon\) of approximately $1.2\times10^{-4}$.

The summary of the pulse parameters for the Transmon design are listed in Table \ref{tab:table5}, which aligns with experimental results reported in \cite{Matekole_2022, McKay_2017}.

\begin{table}[ht]
	\centering
	\caption{Summary of pulse sequencing on the Transmon design.}
	\label{tab:table5}
	\begin{tabular}{c|c|c|c|c|c|c}
		\toprule
		\textbf{Gate} & \textbf{$\Delta\zeta$} & \textbf{$\tau$} & \textbf{$\beta$} & \textbf{$\epsilon$} & \textbf{$\mathcal{F}$}& \textbf{$\mathcal{C}$}\\
		\textbf{} & \text{(rad.)} & \text{(ns)} & \textbf{} & \textbf{($\times 10^{-4}$)} & \text{(\%)}& \text{(gates)}\\
		\midrule
		\textbf{$\pi$} & +0.118 & 28.0 & -0.83 & 4.3 & 99.9779 & 1885 \\
		\text{Hadamard} & +0.109 & 30.0 & -0.85 & 1.2 & 99.9950 & 1759 \\
		\bottomrule
	\end{tabular}
\end{table}

\subsubsection{B.2. Fluxonium benchmarking}

The Fluxonium with transition frequency $\omega_{0 \rightarrow 1}$ of 305.17 MHz is initialized in the state \(\lvert 0 \rangle\) at 20 mK, resulting in a state population of approximately 67.5\%, as determined by Boltzmann statistics. Consequently, \(p_\text{th} = 32.5\%\) of the population is thermally excited to higher energy states during initialization. This corresponds to an excitation rate of \(\Gamma_\uparrow = 212.42\) Hz, reflecting substantial thermal occupation of higher energy levels. Conversely, the relaxation rate is \(\Gamma_\downarrow = 441.18\) Hz, highlighting the Fluxonium's longer effective relaxation time, \(T_1^\text{eff}\).

\begin{figure}[ht]
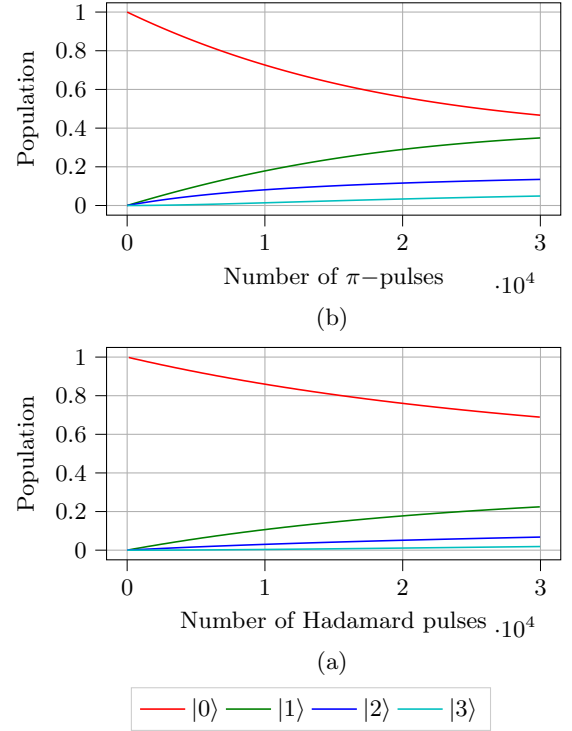

	\centering
	\begin{subfigure}[b]{0.45\textwidth}
		\centering
		\input{Bare-qubit-analytics/FLUX_popdyn_X}
		\label{FLUX_fig_popdyn_pi-pulse}
	\end{subfigure}
	\hfill
	\begin{subfigure}[b]{0.45\textwidth}
		\centering
		\input{Bare-qubit-analytics/FLUX_popdyn_H}
		\label{FLUX_fig_popdyn_H-pulse}
	\end{subfigure}
	\caption{\justifying Randomized benchmarking of the Fluxonium design. (a). The population dynamics of the qubit under a $\pi$-pulse sequence, and (b). a Hadamard gate sequencing.}
	\label{FLUX_popdyn}
\end{figure}

In a manner analogous to the parametric sweep calibration discussed for the Transmon model, the \(\pi\)-pulse duration \(\tau\) for the Fluxonium qubit is optimized to 62.0 ns (\(\delta = 15.5\) ns) with a detuning of \(\Delta\zeta = +0.051\) radians. This optimization achieves an enhanced fidelity of \(\mathcal{F} = 99.997\%\), as indicated by the green circle in Figure~\ref{FLUX_pulse_cal}(a). The error per gate \(\epsilon\), calculated from the randomized benchmarking results shown in Figure~\ref{FLUX_popdyn}(a), is estimated to be \(5.1 \times 10^{-5}\). Consequently, the coherence budget \(\mathcal{C}\) improves significantly to 20,323 gates, attributed primarily to the large \(T_1^\text{eff}\) of the Fluxonium. Similarly, the \(\tau\) for the Hadamard pulse applied to the Fluxonium qubit is optimized to 44.0 ns (\(\delta = 11.0\) ns) with a detuning of \(\Delta\zeta = +0.073\) radians. This optimization yields a fidelity of \(\mathcal{F} = 99.9983\%\), as indicated by the green circle in Figure~\ref{FLUX_pulse_cal}(b). From the randomized benchmarking results shown in Figure~\ref{FLUX_popdyn}(b), the coherence budget \(\mathcal{C}\) is calculated to be 28,637 gates, with the population of the state \(\lvert 0 \rangle\) reduced to 69.74\%. The corresponding error per gate \(\epsilon\) is estimated to be \(3.4 \times 10^{-5}\).

The optimized pulse parameters for the Fluxonium qubit are summarized in Table \ref{tab:table6}, with the values consistent with those experimentally reported results \cite{FastFlux}.

\begin{table}[ht]
	\centering
	\caption{Summary of pulse sequencing on the Fluxonium design.}
	\label{tab:table6}
	\begin{tabular}{c|c|c|c|c|c|c}
		\toprule
		\textbf{Gate} & \textbf{$\Delta\zeta$} & \textbf{$\tau$} & \textbf{$\beta$} & \textbf{$\epsilon$} & \textbf{$\mathcal{F}$}& \textbf{$\mathcal{C}$}\\
		\textbf{} & \text{(rad.)} & \text{(ns)} & \textbf{} & \textbf{($\times 10^{-4}$)} & \text{(\%)}& \text{(gates)}\\
		\midrule
		\textbf{$\pi$} & +0.051 & 62.0 & 0.09 & 5.1 & 99.9966 & 20323 \\
		\text{Hadamard} & +0.073 & 44.0 & 0.09 & 3.4 & 99.9983 & 28637 \\
		\bottomrule
	\end{tabular}
\end{table}

It is important to note that the optimized parameters for the transition from $\lvert 0 \rangle$ to $\lvert + \rangle$ are not directly applicable to the $\lvert 1 \rangle \rightarrow \lvert - \rangle$ transition, due to the different relative phases between the $\lvert 0 \rangle$ and $\lvert 1 \rangle$ states in the $\lvert + \rangle$ and $\lvert - \rangle$ superpositions. However, this is not an issue for the $\pi$-pulse, as it simply flips the Bloch vector along the longitudinal axis.

\subsubsection{B.3. DRAG optimization}

The DRAG optimization, as discussed in Section \hyperref[sec:GateDriving]{II. C.}, is primarily focused on enhancing gate fidelity. This optimization technique is particularly important in weakly anharmonic systems, where leakage errors are minimal for shorter pulse durations but become more prominent as the pulse duration increases \cite{McKay_2017}. Unlike the approach of minimizing leakage errors, the goal of the DRAG method is to reduce the influence of off-resonant transitions that can degrade fidelity \cite{Chen_2016, McKay_2017}. To achieve peak fidelity, the DRAG coefficient $\beta$ is varied parametrically, see Equation (19).

\begin{figure}[ht]
	\centering
	\input{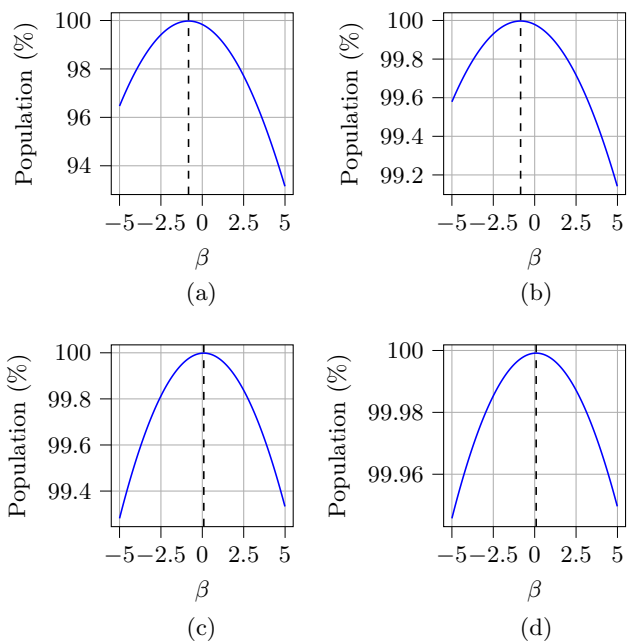}
	\caption{\justifying The DRAG optimization of the optimized gates; (a). $\pi-$pulse and (b). Hadamard pulse on Transmon, (c). $\pi-$pulse and (d). Hadamard pulse on Fluxonium.}
	\label{DRAG}
\end{figure}

For the Transmon model, using the calibrated pulse parameters listed in Table \ref{tab:table5}, the DRAG coefficient $\beta$ is optimized to -0.83 for the $\pi$-pulse and -0.84 for the Hadamard gate, as shown in Figure \ref{DRAG} (a) and \ref{DRAG} (b), respectively. This optimization leads to significant fidelity improvements, with the $\pi$-pulse fidelity increasing from 99.8363\% to 99.9779\%, and the Hadamard gate fidelity rising from 99.9763\% to 99.9950\%.

In contrast, for the Fluxonium model, with the corresponding calibrated pulse parameters given in Table \ref{tab:table6}, the optimal $\beta$ values are determined to be 0.09 for both the $\pi$-pulse and the Hadamard gate, as depicted in Figure \ref{DRAG} (c) and \ref{DRAG} (d). This optimization results in a more modest fidelity enhancement, with the $\pi$-pulse fidelity improving from 99.99625\% to approximately 99.99660\%, and the Hadamard gate fidelity increasing from approximately 99.99833\% to 99.99836\%.

The gate fidelities listed in Table \ref{tab:table5} and Table \ref{tab:table6} represent the final results after the DRAG optimization, demonstrating the improvement in fidelity due to the tuning of the $\beta$ parameter.

\section{IV. Conclusion and Outlook}
\label{sec:concl}

In conclusion, we have developed a detailed framework for the optimization and calibration of superconducting qubit hardware that addresses the spectrum of challenges from design to execution. By integrating circuit design, electromagnetic analysis, spectral analysis, and pulse sequencing, we've created a process that allows for the fine-tuning of qubit parameters to enhance performance. Our application of this framework to Transmon and Fluxonium qubits produced results consistent with earlier experimental observations, reinforcing the validity and practicality of our approach. This work not only advances the methodology for optimizing superconducting qubits but also provides a practical tool for those working toward scalable and fault-tolerant quantum computing systems.
\section{Appendix-A: Bare-Qubit Circuit Analytics}
\label{sec:BareQubits}

This appendix presents the capacitive analysis of the designed qubits. The circuit is enclosed in a Si-vacuum pillbox with a side length of 3 mm. The capacitance matrices were extracted using \texttt{Ansys Q3D Extractor} \cite{ansys_q3d_extractor}, with values converging after 15 iterative passes and error percent of $0.01 \%$ \cite{Squadds}. Referring to Figure \ref{TMON_circuit} (b), the circuit consists of four capacitive pads: the 'Top' and 'Bottom' pads serve as the capacitive islands, connected by a Josephson tunnel junction through which Cooper pairs tunnel. The other two pads, 'Readout' and 'Coupler,' act as capacitive couplings to the bus and islands. In this design, the 'Readout' pad is adjacent to the 'Top' pad, while the 'Coupler' pad is positioned next to the 'Bottom' pad.

\subsection{Transmon circuit design}

The Transmon circuit, with design parameters listed in Table \ref{tab:table1}, comprises qubit islands of dimensions \(400 \, \mu \text{m} \times 125 \, \mu \text{m}\), on a \(\text{SiO}_2\) substrate measuring \(600 \, \mu \text{m} \times 600 \, \mu \text{m}\). The island dimensions are chosen to ensure a low charging energy, \(E_C\), and a high \(E_J/E_C\) ratio, achieved through the inverse relationship between capacitance and the metal pad area.

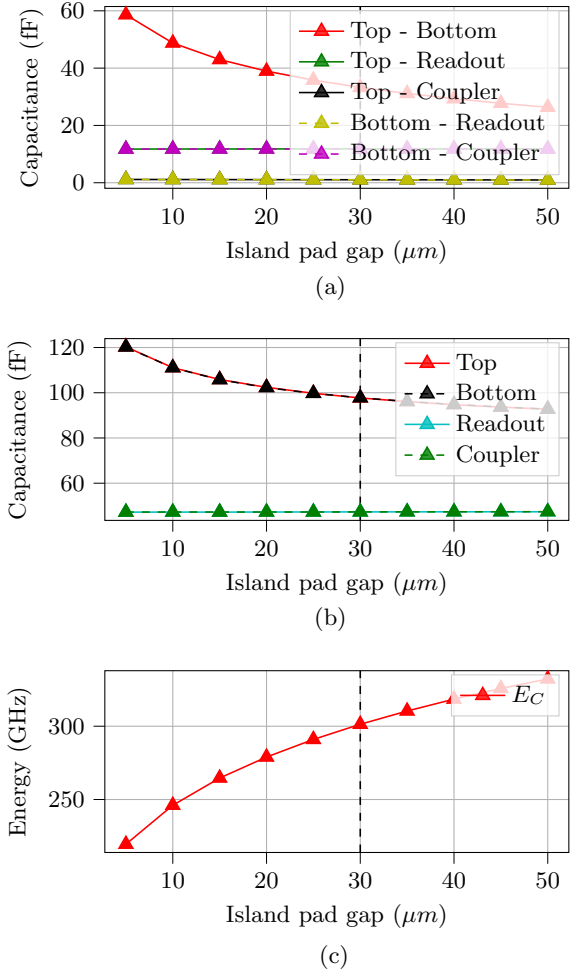
\begin{figure}[ht]
	\renewcommand{\thefigure}{A\arabic{figure}}
	\setcounter{figure}{0}
	\centering
\begin{tikzpicture}

\definecolor{darkgray176}{RGB}{176,176,176}
\definecolor{darkturquoise0191191}{RGB}{0,191,191}
\definecolor{darkviolet1910191}{RGB}{191,0,191}
\definecolor{goldenrod1911910}{RGB}{191,191,0}
\definecolor{green01270}{RGB}{0,127,0}
\definecolor{lightgray204}{RGB}{204,204,204}

\begin{groupplot}[group style={group size=1 by 3, vertical sep=2cm}, width=7.75cm, height=4cm]
\nextgroupplot[
legend cell align={left},
legend style={
  fill opacity=0.8,
  draw opacity=1,
  text opacity=1,
  draw=lightgray204
},
tick align=outside,
tick pos=left,
x grid style={darkgray176},
xlabel={Island pad gap (\(\displaystyle \mu m\))},
xmajorgrids,
xmin=2.75, xmax=52.25,
xtick style={color=black},
y grid style={darkgray176},
ylabel={Capacitance (fF)},
ymajorgrids,
ymin=-1.936595, ymax=61.496115,
ytick style={color=black}
]
\addplot [semithick, red, mark=triangle*, mark size=3, mark options={solid}]
table {%
5 58.61281
10 48.75308
15 42.98424
20 38.95053
25 35.82368
30 33.27775
35 31.15428
40 29.32514
45 27.72929
50 26.33037
};
\addlegendentry{Top - Bottom}
\addplot [semithick, green01270, mark=triangle*, mark size=3, mark options={solid}]
table {%
5 11.74236
10 11.77113
15 11.77816
20 11.79786
25 11.80882
30 11.80774
35 11.81148
40 11.81145
45 11.81085
50 11.80568
};
\addlegendentry{Top - Readout}
\addplot [semithick, black, mark=triangle*, mark size=3, mark options={solid}]
table {%
5 1.1407
10 1.12165
15 1.10018
20 1.07979
25 1.05856
30 1.03859
35 1.01887
40 0.99882
45 0.97948
50 0.96102
};
\addlegendentry{Top - Coupler}
\addplot [semithick, goldenrod1911910, dashed, mark=triangle*, mark size=3, mark options={solid}]
table {%
5 1.13431
10 1.1144
15 1.09124
20 1.07024
25 1.04877
30 1.02851
35 1.00841
40 0.98804
45 0.96781
50 0.94671
};
\addlegendentry{Bottom - Readout}
\addplot [semithick, darkviolet1910191, dashed, mark=triangle*, mark size=3, mark options={solid}]
table {%
5 11.73262
10 11.75791
15 11.75557
20 11.76757
25 11.77375
30 11.78363
35 11.79116
40 11.79252
45 11.79795
50 11.7893
};
\addlegendentry{Bottom - Coupler}
\addplot [semithick, black, dashed, forget plot]
table {%
30 -1.93659500000001
30 61.496115
};

\nextgroupplot[
legend cell align={left},
legend style={
  fill opacity=0.8,
  draw opacity=1,
  text opacity=1,
  draw=lightgray204
},
tick align=outside,
tick pos=left,
x grid style={darkgray176},
xlabel={Island pad gap (\(\displaystyle \mu m\))},
xmajorgrids,
xmin=2.75, xmax=52.25,
xtick style={color=black},
y grid style={darkgray176},
ylabel={Capacitance (fF)},
ymajorgrids,
ymin=43.555735, ymax=123.942965,
ytick style={color=black}
]
\addplot [semithick, red, mark=triangle*, mark size=3, mark options={solid}]
table {%
5 120.289
10 111.092
15 105.854
20 102.407
25 99.7657
30 97.6919
35 96.0915
40 94.7498
45 93.6239
50 92.738
};
\addlegendentry{Top}
\addplot [semithick, black, dashed, mark=triangle*, mark size=3, mark options={solid}]
table {%
5 120.2268
10 111.054
15 105.7802
20 102.2829
25 99.72771
30 97.77479
35 96.15821
40 94.80317
45 93.71765
50 92.76781
};
\addlegendentry{Bottom}
\addplot [semithick, darkturquoise0191191, mark=triangle*, mark size=3, mark options={solid}]
table {%
5 47.2097
10 47.2314
15 47.21363
20 47.25634
25 47.28457
30 47.30608
35 47.33057
40 47.34585
45 47.35722
50 47.35709
};
\addlegendentry{Readout}
\addplot [semithick, green01270, dashed, mark=triangle*, mark size=3, mark options={solid}]
table {%
5 47.21245
10 47.25822
15 47.24471
20 47.25998
25 47.25309
30 47.29456
35 47.32172
40 47.34819
45 47.37093
50 47.40597
};
\addlegendentry{Coupler}
\addplot [semithick, black, dashed, forget plot]
table {%
30 43.555735
30 123.942965
};

\nextgroupplot[
legend cell align={left},
legend style={
  fill opacity=0.8,
  draw opacity=1,
  text opacity=1,
  draw=lightgray204
},
tick align=outside,
tick pos=left,
x grid style={darkgray176},
xlabel={Island pad gap (\(\displaystyle \mu m\))},
xmajorgrids,
xmin=2.75, xmax=52.25,
xtick style={color=black},
y grid style={darkgray176},
ylabel={Energy (GHz)},
ymajorgrids,
ymin=213.8621, ymax=337.8739,
ytick style={color=black}
]
\addplot [semithick, red, mark=triangle*, mark size=3, mark options={solid}]
table {%
5 219.499
10 246.079
15 264.631
20 278.963
25 290.986
30 301.327
35 310.361
40 318.49
45 325.751
50 332.237
};
\addlegendentry{$E_C$}
\addplot [semithick, black, dashed, forget plot]
table {%
30 213.8621
30 337.8739
};
\end{groupplot}
\node[above] at (3, -1.6) {(a)};
\node[above] at (3, -6) {(b)};
\node[above] at (3.05, -10.5) {(c)};
\end{tikzpicture}
	\caption{\justifying Capacitance analysis of the Transmon circuit with varying island pad gap: (a) cross-capacitance between pads, (b) self-capacitance of pads, and (c) effective capacitive energy, $E_C$, as a function of island pad gap.}
	\label{TMON_fig_topology}
\end{figure}

Figure \ref{TMON_fig_topology} illustrates the capacitive properties of the Transmon circuit as the island pad gap is varied from $5 \, \mu\text{m}$ to $50 \, \mu\text{m}$, while maintaining the bus pad gaps relative to their respective islands. Figure \ref{TMON_fig_topology} (a) shows a decrease in the inter-island capacitance with increasing gap, indicating a higher effective $E_C$ between the top and bottom pads. The bus pads exhibit negligible dependence on the pad gap due to constant capacitive coupling with their corresponding islands.

Figure \ref{TMON_fig_topology}(b) shows a reduction in the self-capacitance of the island pads with increased gap, leading to an overall increase in the qubit's effective capacitive energy, $E_C$, as shown in Figure \ref{TMON_fig_topology}(c). An island gap of $30 \, \mu \text{m}$ is selected to maintain a low $E_C$ and a high $E_J/E_C$ ratio. A higher $E_J$ would reduce anharmonicity, raising the risk of leakage during pulse sequences.

\begin{figure}[ht]
	\renewcommand{\thefigure}{A\arabic{figure}}
	\centering
	\input{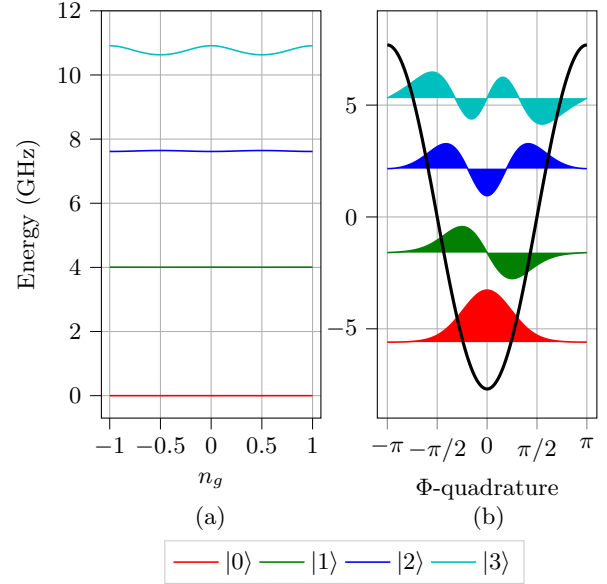}
	\caption{\justifying The Transmon design's (a). energy as a function of $n_g$, and (b). wavefunctions in the $\Phi$ quadrature.}
	\label{TMON_fig_wavefunc}
\end{figure}

The parameters discussed in Section \hyperref[sec:singlequbits]{III. A.1.}, Table \ref{tab:table2} yield a Transmon spectrum as shown in Figure \ref{TMON_fig_wavefunc}. Figure \ref{TMON_fig_wavefunc} (a) demonstrates the weak anharmonicity of the qubit, indicated by the energy gap between \(\lvert 0 \rangle \to \lvert 1 \rangle\) and \(\lvert 1 \rangle \to \lvert 2 \rangle\), as also demonstrated by the spectral dispersions in Figure \ref{TMON_fig_specs} (d). The qubit transitions are largely independent of \(n_g\), owing to minimized spectral dispersions. Figure \ref{TMON_fig_wavefunc} (b) illustrates harmonic-like wavefunctions within the quadratic approximation of the potential for the qubit subspace, with anharmonicity manifesting in the higher levels as the potential tapers.

\subsection{Fluxonium circuit design}

The Fluxonium qubit islands, with design parameters listed in Table \ref{tab:table3}, have dimensions of \(250 \, \mu \text{m} \times 45 \, \mu \text{m}\) on a \(\text{SiO}_2\) substrate measuring \(600 \, \mu \text{m} \times 600 \, \mu \text{m}\). The island dimensions are selected to ensure a high charging energy, \(E_C\).

 \begin{figure}[ht]
 	\renewcommand{\thefigure}{A\arabic{figure}}
	\centering
\begin{tikzpicture}

\definecolor{darkgray176}{RGB}{176,176,176}
\definecolor{darkturquoise0191191}{RGB}{0,191,191}
\definecolor{darkviolet1910191}{RGB}{191,0,191}
\definecolor{goldenrod1911910}{RGB}{191,191,0}
\definecolor{green01270}{RGB}{0,127,0}
\definecolor{lightgray204}{RGB}{204,204,204}

\begin{groupplot}[group style={group size=1 by 3, vertical sep=2cm}, width=7.75cm, height=4cm]
\nextgroupplot[
legend cell align={left},
legend style={
  fill opacity=0.8,
  draw opacity=1,
  text opacity=1,
  draw=lightgray204
},
tick align=outside,
tick pos=left,
x grid style={darkgray176},
xlabel={Island pad gap (\(\displaystyle \mu m\))},
xmajorgrids,
xmin=11.25, xmax=313.75,
xtick style={color=black},
y grid style={darkgray176},
ylabel={Capacitance (fF)},
ymajorgrids,
ymin=-0.248127, ymax=16.547927,
ytick style={color=black}
]
\addplot [semithick, red, mark=triangle*, mark size=3, mark options={solid}]
table {%
25 15.78447
50 10.71555
75 8.06097
100 6.4038
125 5.25411
150 4.39476
175 3.71726
200 3.1655
225 2.71048
250 2.34568
275 2.0458
300 1.79514
};
\addlegendentry{Top - Bottom}
\addplot [semithick, green01270, mark=triangle*, mark size=3, mark options={solid}]
table {%
25 10.41473
50 10.67034
75 10.80501
100 10.88692
125 10.93167
150 10.93573
175 10.92843
200 10.88129
225 10.8
250 10.70659
275 10.58188
300 10.4198
};
\addlegendentry{Top - Readout}
\addplot [semithick, black, mark=triangle*, mark size=3, mark options={solid}]
table {%
25 2.11921
50 1.87049
75 1.65332
100 1.46596
125 1.29276
150 1.13515
175 1.00391
200 0.89258
225 0.78795
250 0.69118
275 0.59873
300 0.51533
};
\addlegendentry{Top - Coupler}
\addplot [semithick, goldenrod1911910, dashed, mark=triangle*, mark size=3, mark options={solid}]
table {%
25 2.13873
50 1.90375
75 1.69802
100 1.49213
125 1.28731
150 1.12824
175 0.99229
200 0.877
225 0.77656
250 0.6835
275 0.59998
300 0.52649
};
\addlegendentry{Bottom - Readout}
\addplot [semithick, darkviolet1910191, dashed, mark=triangle*, mark size=3, mark options={solid}]
table {%
25 10.40922
50 10.66593
75 10.824
100 10.90125
125 10.92985
150 10.93394
175 10.91448
200 10.86929
225 10.81205
250 10.71781
275 10.5839
300 10.42086
};
\addlegendentry{Bottom - Coupler}
\addplot [semithick, black, dashed, forget plot]
table {%
270 -0.248126999999997
270 16.547927
};

\nextgroupplot[
legend cell align={left},
legend style={
  fill opacity=0.8,
  draw opacity=1,
  text opacity=1,
  draw=lightgray204
},
tick align=outside,
tick pos=left,
x grid style={darkgray176},
xlabel={Island pad gap (\(\displaystyle \mu m\))},
xmajorgrids,
xmin=11.25, xmax=313.75,
xtick style={color=black},
y grid style={darkgray176},
ylabel={Capacitance (fF)},
ymajorgrids,
ymin=40.2553495, ymax=54.7280605,
ytick style={color=black}
]
\addplot [semithick, red, mark=triangle*, mark size=3, mark options={solid}]
table {%
25 47.4168
50 43.9283
75 42.5363
100 41.7947
125 41.3378
150 41.1268
175 41.0069
200 40.915
225 40.9132
250 40.9657
275 41.0073
300 41.1752
};
\addlegendentry{Top}
\addplot [semithick, black, dashed, mark=triangle*, mark size=3, mark options={solid}]
table {%
25 47.35768
50 43.94918
75 42.53001
100 41.75893
125 41.36554
150 41.13938
175 40.96412
200 40.92052
225 40.93503
250 40.94055
275 41.04492
300 41.19175
};
\addlegendentry{Bottom}
\addplot [semithick, darkturquoise0191191, mark=triangle*, mark size=3, mark options={solid}]
table {%
25 51.62804
50 51.66081
75 51.67734
100 51.72223
125 51.80774
150 51.87477
175 51.98851
200 52.17191
225 52.38059
250 52.68427
275 53.2097
300 54.07021
};
\addlegendentry{Readout}
\addplot [semithick, green01270, dashed, mark=triangle*, mark size=3, mark options={solid}]
table {%
25 51.66022
50 51.66433
75 51.66769
100 51.73258
125 51.78958
150 51.86516
175 52.02021
200 52.16771
225 52.37058
250 52.70751
275 53.18892
300 54.04975
};
\addlegendentry{Coupler}
\addplot [semithick, black, dashed, forget plot]
table {%
270 40.2553495
270 54.7280605
};

\nextgroupplot[
legend cell align={left},
legend style={
  fill opacity=0.8,
  draw opacity=1,
  text opacity=1,
  draw=lightgray204
},
tick align=outside,
tick pos=left,
x grid style={darkgray176},
xlabel={Island pad gap (\(\displaystyle \mu m\))},
xmajorgrids,
xmin=11.25, xmax=313.75,
xtick style={color=black},
y grid style={darkgray176},
ylabel={Energy (GHz)},
ymajorgrids,
ymin=610.57065, ymax=956.70435,
ytick style={color=black}
]
\addplot [semithick, red, mark=triangle*, mark size=3, mark options={solid}]
table {%
25 626.304
50 728.558
75 790.785
100 833.514
125 864.309
150 886.799
175 904.753
200 918.445
225 928.49
250 935.79
275 940.204
300 940.971
};
\addlegendentry{$E_C$}
\addplot [semithick, black, dashed, forget plot]
table {%
270 610.57065
270 956.70435
};
\end{groupplot}

\node[above] at (3, -1.6) {(a)};
\node[above] at (3, -6) {(b)};
\node[above] at (3.05, -10.5) {(c)};

\end{tikzpicture}
	\caption{\justifying Capacitance analysis of the Fluxonium circuit with varying island pad gap: (a) cross-capacitance between pads, (b) self-capacitance of pads, and (c) effective capacitive energy, $E_C$, as a function of island pad gap.}
	\label{FLUX_fig_topology}
\end{figure}
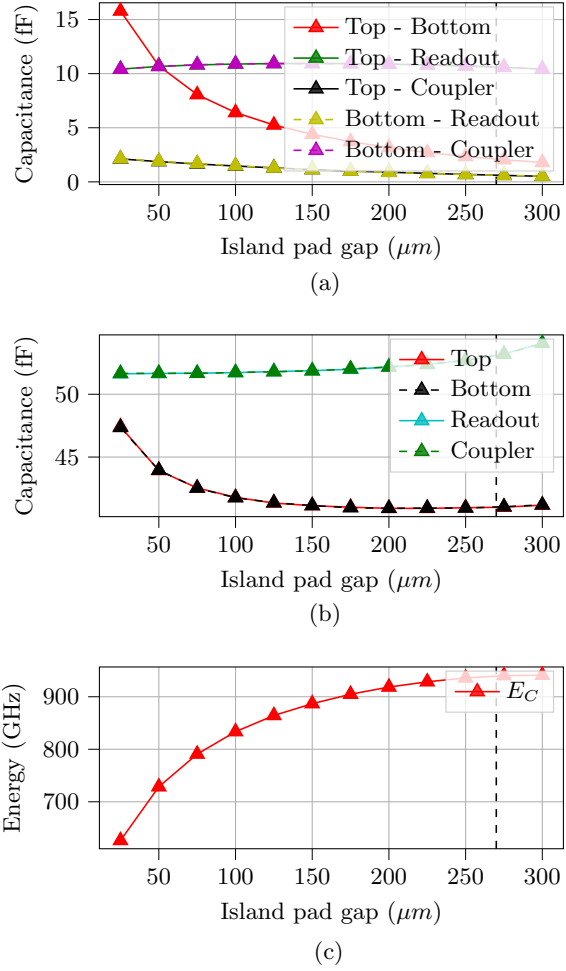
 
Figure \ref{FLUX_fig_topology} illustrates the capacitive properties of the Fluxonium circuit as the island pad gap is varied from $230 \, \mu\text{m}$ to $270 \, \mu\text{m}$, while maintaining the bus pad gaps relative to their respective islands. Figure \ref{FLUX_fig_topology}(a) shows a slowly descending island cross-capacitance with increasing gap. Figure \ref{FLUX_fig_topology}(b) shows the variation in self-capacitances of the islands. A higher island gap of $270 \, \mu \text{m}$ is selected to maintain a high $E_C$, as seen in Figure \ref{FLUX_fig_topology}(c)
 
\begin{figure}[ht]
	\renewcommand{\thefigure}{A\arabic{figure}}
    \centering
    \input{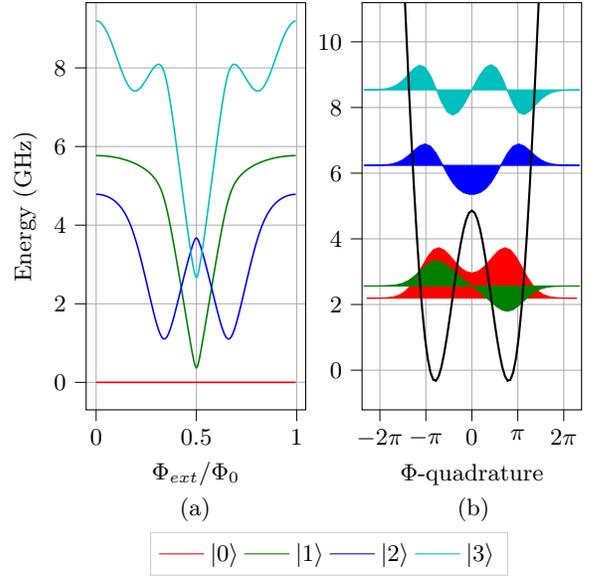}
    \caption{\justifying The Fluxonium qubit's: (a) transmission spectrum, where the $\lvert 0 \rangle$ and $\lvert 1 \rangle$ states are closely spaced at the flux-frustration point; (b) the wavefunctions, with the computational states localized within the symmetric double-well potential around the flux-frustration point.}
    \label{FLUX_fig_wavefunc}
    \addtocounter{figure}{-1}
    \renewcommand{\thefigure}{\arabic{figure}}
\end{figure}

The parameters discussed in Section \hyperref[sec:singlequbits]{III. A.2.} and listed in Table \ref{tab:table4} yield a Fluxonium spectrum as shown in Figure \ref{FLUX_fig_wavefunc}. Figure \ref{FLUX_fig_wavefunc} (a) illustrates the spectrum, highlighting the advantages of a low transition frequency and high anharmonicity at the half-integer flux quantum. Figure \ref{FLUX_fig_wavefunc} (b) depicts the wavefunctions of the qubit, with the qubit subspace closely spaced. The fictitious particle in this double-well potential tunnels across the wells within the qubit subspace, with its tunneling rate controlled by the \(E_J/E_C\) ratio \cite{21}.
\section{Appendix-B: Decoherence Analysis}
\label{sec:BareQubitsCoherence}

The following section discusses the noise channels considered for estimating the effective coherence times of the Transmon and Fluxonium models. The theoretical background on coherence times is presented in Section \hyperref[sec:coherencetimes]{II.C}.

\subsection{B.1 Transmon}

The effective longitudinal relaxation rates of the Transmon are primarily determined by dielectric losses at the AlOx layer in the Josephson junction, as illustrated in Figure~\ref{TMON_fig_relaxations} (a). This analysis employs the PSD from \cite{32}, which assumes a dielectric quality factor $Q_{cap} = 1.3 \times 10^6$, with the corresponding noise operator for this channel being $2e\hat{N}$.

\begin{figure}[ht]
	\renewcommand{\thefigure}{B\arabic{figure}}
	\setcounter{figure}{0}
	\centering
	\input{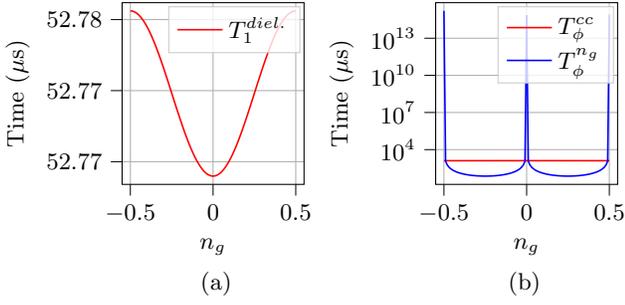}
	\caption{\justifying (a). Longitudinal relaxations due to dielectric losses at the Josephson junction, and (b). pure-dephasing mainly due to critical current noise at the Josephson junction of the Transmon.}
	\label{TMON_fig_relaxations}
\end{figure}

The pure dephasing of the Transmon model is illustrated in Figure~\ref{TMON_fig_relaxations} (b), indicating the insensitivity to charge noise at the external gate when operated at half-integer values of $n_g$. The qubit dephases primarily due to charge noise and critical current noise from defects in the AlOx layer of the Josephson junction. The main contribution to dephasing arises from fluctuations in charge densities at the oxide-junction caused by defect sites, which affect the tunneling rate across the Josephson junction and, consequently, the critical current of the device. The noise strength for this source is set to $10^{-7} I_C$, with the corresponding noise operator given by $\partial \hat{H} / \partial I_c$ \cite{9}.

\subsection{B.2 Fluxonium}

\begin{figure}[ht]
	\renewcommand{\thefigure}{B\arabic{figure}}
	\centering
	\input{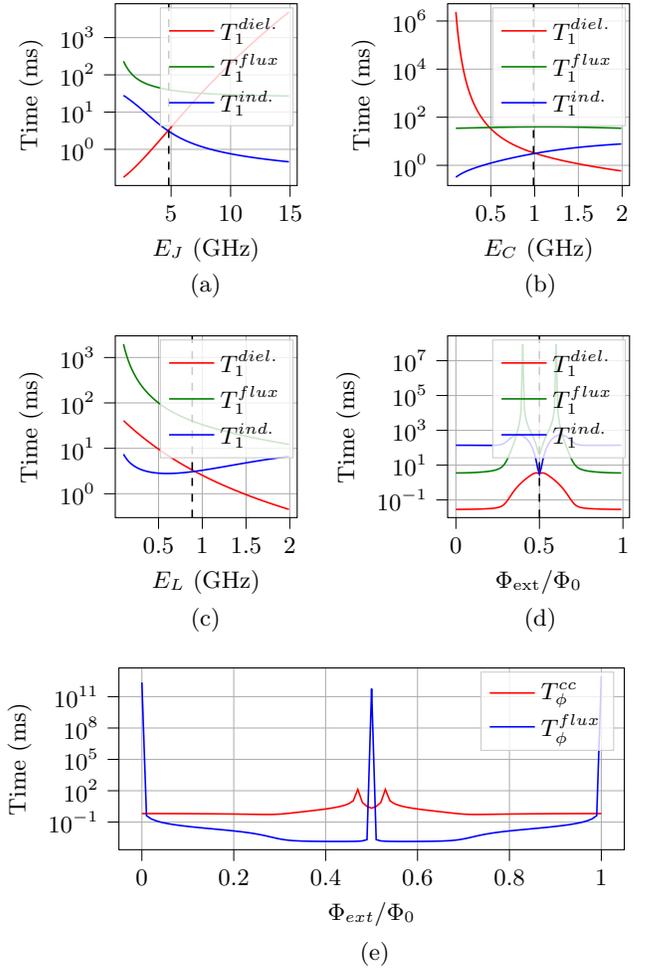}
	\caption{\justifying Longitudinal relaxation of the Fluxonium qubit due to dielectric noise, flux-bias noise, and quasiparticle tunneling: (a) versus $E_J$, (b) versus $E_C$, (c) versus $E_L$, with optimized values marked by a dashed line. (d) Coherence times peak symmetrically around the flux-frustration point. (e) $1/f$ dephasing due to critical current and flux-bias noise.}
	\label{FLUX_fig_relaxations}
\end{figure}

The effective longitudinal relaxation rates of the Fluxonium qubit arise from: (a) dielectric losses at the Josephson junction of the qubit loop, (b) decoherence due to flux-bias noise across the bias loop, and (c) quasiparticle tunneling through the Josephson junction chain in the superinductor. This is illustrated in Figure \ref{FLUX_fig_relaxations} (a), \ref{FLUX_fig_relaxations} (b), and \ref{FLUX_fig_relaxations} (c), for sweeping \(E_J\), \(E_C\), and \(E_L\), respectively. The flux noise remains largely invariant with respect to \(E_C\). The optimized design values for Fluxonium, as shown in Table \ref{tab:table4}, indicate that \(T_1^{\text{diel.}} \simeq T_1^{\text{ind.}}\) and \(T_1^{\text{flux}} > T_1^{\text{diel.}}, T_1^{\text{ind.}}\). For flux noise, we use the PSD from \cite{33}, with a bias line impedance of \(50 \, \Omega\) and a mutual inductance of \(400 \, \Phi_0 \, \text{A}^{-1}\). The associated noise operator is \(\partial \hat{H} / \partial \Phi_{\text{ext}}\) \cite{17, Bylander_2011}.

The inductive noise arising from quasiparticles (i.e., unpaired electrons, with a density ranging from \(10^{-6}\) to \(10^{-8}\) per Cooper pair for temperatures $\le 150$ mK \cite{Krantz2019}) tunneling across a superinductor is modeled using the power spectral density (PSD) described in \cite{32}. This noise is characterized by an inductive quality factor of \(Q_{\text{ind}} = 392.0 \times 10^6\), with the corresponding noise operator given as \(\Phi_0 \hat{\Phi}\).

%
%

\renewcommand{\bibname}{References}

\bibliography{references}

\end{document}